\documentclass[a4paper]{cas-dc}

\usepackage[authoryear]{natbib}
\usepackage{hyperref}
\usepackage{graphicx}
\usepackage{subfigure}
\usepackage{epstopdf}
\usepackage{color}
\usepackage[noend]{algpseudocode}
\usepackage{algorithmicx}
\usepackage{cuted}
\usepackage{algorithm}
\usepackage{dsfont}

\usepackage{soul}
\usepackage{color, xcolor}
\usepackage[T1]{fontenc}

\usepackage{booktabs}
\usepackage{multirow}
\usepackage{amssymb}
\usepackage{pifont} % This package is needed for the \ding{55} symbol
\usepackage{tcolorbox}
\usepackage{colortbl}

\usepackage{booktabs}
\usepackage{array, caption, threeparttable}
\usepackage[font=small,labelfont=bf,labelsep=none]{caption}
\def\tsc#1{\csdef{#1}{\textsc{\lowercase{#1}}\xspace}}
\tsc{WGM}
\tsc{QE}
\tsc{EP}
\tsc{PMS}
\tsc{BEC}
\tsc{DE}
\begin{document}
\begin{sloppypar}
	\let\WriteBookmarks\relax
	\def\floatpagepagefraction{1}
	\def\textpagefraction{.001}
	\let\printorcid\relax
	\shorttitle{}
	\shortauthors{C. Xu et~al.} %% 缩略作者 自己名字， 比如： 张三 = S. Zhang

	%% 标题
	\title [mode = title]{UMind: A Unified Multitask Network for Zero-Shot M/EEG Visual Decoding}
	%%\tnotemark[1,2]

	%%\tnotetext[1]{This document is the results of the research project funded by the National Science Foundation.}

	%%\tnotetext[2]{The second title footnote which is a longer text matter to fill through the whole text width and overflow into another line in the footnotes area of the first page.}

	%% 作者顺序
	%% 1
	\author[1, 2]{\textcolor[RGB]{0,0,1}{Chengjian Xu}}
	% \cormark[1]%%通讯作者星标
	% \ead{cj.xu@siat.ac.cn}
        \address[1]{Shenzhen Institutes of Advanced Technology, Chinese Academy of Sciences, China}
        \address[2]{University of Chinese Academy of Sciences, China}

	%% 2
	\author[3]{\textcolor[RGB]{0,0,1}{Yonghao Song}}
        \address[3]{Department of Biomedical Engineering, Tsinghua University, China}
        % \ead{songyh22@mails.tsinghua.edu.cn}

        %% 3
	\author[4]{\textcolor[RGB]{0,0,1}{Zelin Liao}}
        \address[4]{School of Automation, Guangdong University of Technology, China}
        % \ead{2112404424@mail2.gdut.edu.cn}

        %% 4
	\author[4]{\textcolor[RGB]{0,0,1}{Haochuan Zhang}}
        % \ead{haochuan.zhang@gdut.edu.cn}
        
        %% 5
	\author[5]{\textcolor[RGB]{0,0,1}{Qiong Wang}}
        \address[5]{Guangdong Provincial Key Laboratory of Computer Vision and Virtual Reality Technology, Shenzhen Institutes of Advanced Technology, China}
        % \ead{wangqiong@siat.ac.cn}

        %% 6
        \author[6]{\textcolor[RGB]{0,0,1}{Qingqing Zheng}}
        \address[6]{
        Artificial Intelligence Research Institute, Shenzhen University of Advanced Technology, China}
	% \fnmark[3] %%第几作者
	\cormark[1]%%通讯作者星标
	\ead{zhengqingqing@suat-sz.edu.cn}

	\cortext[cor1]{Corresponding author.} %% 首页左下角通讯作者
	%%\cortext[cor2]{Principal corresponding author} 
        %% 共一
	% \fntext[fn1]{Equal Contribution.}
	%%\fntext[fn2]{Another author footnote, this is a very long footnote and it should be a really long footnote. But this footnote is not yet sufficiently long enough to make two lines of footnote text.}

	%%\nonumnote{This note has no numbers. In this work we demonstrate $a_b$ the formation Y\_1 of a new type of polariton on the interface between a cuprous oxide slab and a polystyrene micro-sphere placed on the slab.}

	%%摘要
	\begin{abstract}
        Decoding visual information from time-resolved brain recordings, such as EEG and MEG, plays a pivotal role in real-time brain-computer interfaces.
        However, existing approaches primarily focus on direct brain-image feature alignment and are limited to single-task frameworks or task-specific models.
        In this paper, we propose a \textbf{U}nified \textbf{M}ult\textbf{I}task \textbf{N}etwork for zero-shot M/EEG visual \textbf{D}ecoding (referred to \textbf{UMind}), including visual stimulus retrieval, classification, and reconstruction, where multiple tasks mutually enhance each other.
        Our method learns robust neural-visual and semantic representations through multimodal alignment with both image and text modalities.
        The integration of both coarse and fine-grained texts enhances the extraction of these neural representations, enabling more detailed semantic and visual decoding.
        These representations then serve as dual conditional inputs to a pre-trained diffusion model, guiding visual reconstruction from both visual and semantic perspectives.
        Extensive evaluations on MEG and EEG datasets demonstrate the effectiveness, robustness, and biological plausibility of our approach in capturing spatiotemporal neural dynamics. 
        Our approach sets a multitask pipeline for brain visual decoding, highlighting the synergy of semantic information in visual feature extraction. The code is available at \href{https://github.com/xuchengjian632/UMind}{https://github.com/xuchengjian632/UMind}.
	\end{abstract}

	% 	%%\includegraphics{./grabs.pdf} %%图片摘要地址路径

	%%高亮
	% \begin{highlights}
	% 	\item End-to-end community detection method based on graph convolution network.
	% 	\item A new community perspective similarity is proposed.
	% 	\item Modify the convolution layer for large networks.
	% 	\item The loss function based on modularity and Bernoulli Poisson model is introduced.
	% 	\item Evaluate performance using real-world networks.
	% \end{highlights}
		
	%% 关键词

	% 此指令为生成标题格式，不可删除
	\maketitle

	%% 2.引言
	\section{Introduction}
        \label{sec:Introduction}
% Decoding and reconstructing visual stimuli from brain activity signals is a challenging yet highly significant task in neuroscience.
% A major challenge in neuroscience is the objective assessment of visual perceptual experience by decoding mental content from brain activity~\citep{miyawakiVisualImageReconstruction2008,kayIdentifyingNaturalImages2008,jiangBrainmediaDeepFramework2020}.
% With the rapid advancements in brain-computer interfaces (BCIs) and image generation technologies, it has become increasingly possible to explore the brain's representation of visual information.
In the realm of brain-computer interfaces (BCIs), visual information decoding from neural signals is of paramount importance in understanding the intricate processes of visual perception and cognition~\citep{miyawakiVisualImageReconstruction2008}.
It forms the foundation for advancing our comprehension of how the brain interprets and processes visual stimuli, including tasks such as \textit{visual classification}, \textit{retrieval}, and \textit{reconstruction}.
% Decoding and reconstructing brain activity related to visual perception not only offers insights into brain function but also holds great potential for developing assistive technologies for individuals with sensory impairments.
Visual decoding has wide-ranging applications, offering significant potential to provide alternative communication and control mechanisms, thereby enhancing the autonomy and capabilities of individuals with severe motor impairments~\citep{kayIdentifyingNaturalImages2008}.
As non-invasive brain imaging technologies, electroencephalography (EEG) and magnetoencephalography (MEG) have shown great potential in the field of visual stimulus decoding~\citep{songDecodingNaturalImages2024,wang2025eegmamba,borra2025protocol}.

\begin{comment}
Benefiting from the high spatial resolution of functional magnetic resonance imaging (fMRI) signals, previous studies have predominantly focused on decoding and reconstructing visual images from extensive fMRI data~\citep{allenMassive7TFMRI2022,linMindReaderReconstructing2022,takagiHighresolutionImageReconstruction2023}.
However, blood oxygen level-dependent (BOLD) signals measured via fMRI are very slow, so it has been difficult to apply fMRI for real-time decoding of brain activity during visual stimuli.
In addition, the practical application of fMRI is constrained by its high cost, large volume of data, and the complexity involved in data processing~\citep{fuBrainVisExploringBridge2024}.
By contrast, electroencephalography (EEG) signals, with higher temporal resolution, lower cost, and ease of processing, are better suited for real-time decoding~\citep{congLinkingBrainResponses2013,liuEEGbasedStudyPerception2019} of brain activity in response to visual stimuli.
Similarly, magnetoencephalography (MEG) signals, with characteristics similar to those of EEG signals, have also been used for visual stimulus decoding and reconstruction~\citep{benchetritBrainDecodingRealtime2024}.
Additionally, MEG offers higher spatial resolution compared to EEG, making it a valuable source of guiding evidence for the cost-effective application of EEG.
\end{comment}

It is noteworthy that learning robust visual representations from M/EEG signals for visual stimulus decoding and reconstruction has become a prominent research focus~\citep{spampinatoDeepLearningHuman2017,jiaoDecodingEEGVisualguided2019,liVisualDecodingReconstruction2024}.
Although numerous studies have focused on decoding M/EEG signals, most existing methods are designed for single-task scenarios, such as visual stimulus reconstruction or classification. 
For instance, 
Song \textit{et al.}~\citep{songDecodingNaturalImages2024} introduced a simple but efficient contrastive learning framework NICE, which aligns EEG and image representations for decoding images from EEG signals.
The NICE demonstrated the feasibility of feature alignment and its biological plausibility.
Similarly, Benchetrit \textit{et al.}~\citep{benchetritBrainDecodingRealtime2024} firstly aligns MEG features with CLIP image embeddings for retrieval and reconstruction of real-time visual stimuli.
% However, these methods are limited to single-task scenarios and fail to propose a unified multitask framework that enables mutual reinforcement among different tasks.
However, these methods remain confined to single-task objectives, lacking a unified framework to mutually reinforce multiple tasks (e.g., retrieval, classification, and reconstruction). 
% They primarily focus on aligning M/EEG signals with image representations, neglecting the integration of textual semantic information.
Furthermore, they primarily focus on aligning M/EEG signals with visual embeddings, while neglecting the integration of textual semantic information, thereby limiting their ability to utilize rich contextual semantics for accurate cross-modal understanding and inference.
% Consequently, they do not take advantage of the rich semantic context provided by text for a more comprehensive understanding of visual stimuli, and thus their ability to perform accurate retrieval and classification is limited.

Some studies have attempted to design multitask frameworks for zero-shot visual stimulus decoding. 
% Li \textit{et al.}~\citep{liVisualDecodingReconstruction2024} proposed a novel brain decoding framework that enables zero-shot visual stimulus classification, retrieval, and reconstruction using EEG and MEG data.
A representative example is the work by Li et al.~\citep{liVisualDecodingReconstruction2024}, which proposes a novel brain decoding framework that enables zero-shot visual stimulus classification, retrieval, and reconstruction using EEG and MEG data.
The method achieves results comparable to visual stimulus decoding using fMRI data with a customized M/EEG encoder and a two-stage M/EEG-to-image generation strategy.
Although this work presents a multitask model, its underlying implementations still rely on separate M/EEG encoders for each task, 
% which prevents the model from leveraging shared information across tasks, thereby hindering the mutual reinforcement and collaboration among them.
thereby preventing cross-task knowledge sharing and mutual reinforcement. 
Furthermore, while the framework incorporates coarse-grained textual categories (e.g., "cat" or "raspberry") for visual stimulus reconstruction, it fails to perform multimodal alignment between M/EEG signals, images, and text, 
resulting in generated images or retrieval outcomes with compromised semantic consistency and visual fidelity. 
% which could help the model better focus on the semantic representations within visual information.
Most critically, it fails to utilize fine-grained textual descriptions (e.g., “the aircraft
carrier is sailing in the ocean”), 
which fundamentally constrains their applicability in complex scenarios requiring high-precision representations, such as fine-grained image generation or semantically ambiguous stimulus retrieval. 
% it does not fully utilize fine-grained textual descriptions of visual stimuli to assist the model in extracting M/EEG representations enriched with more detailed semantic information.

% However, this work does not employ a unified multitask framework.
% Instead, it trains a separate EEG encoder for each task.
% Nevertheless, these works do not present a unified model capable of handling multiple tasks while allowing them to mutually enhance each other to achieve higher-quality representations.
% Moreover, they have overlooked the incorporation of text modality, which could help focus more effectively on the semantic representations within visual information.

To address the above challenges, we propose \textbf{UMind}, a unified multitask framework for zero-shot visual stimulus retrieval, classification, and reconstruction. 
This framework learns robust neural visual/semantic representations through cross-modal alignment with visual stimuli and textural semantic information.
% To further enhance the extraction of visual representations while capturing more detailed and critical semantic representations from M/EEG signals, we incorporate text modality, including coarse-grained text with only category information and fine-grained text providing detailed descriptions of images for multimodal alignment.
% The coarse-grained text contains only categorical information and lacks detailed descriptions of visual stimuli, such as color, shape, and orientation.
% In contrast, the fine-grained text generated by pre-trained large multimodal models (LMMs) includes richer details but may provide ambiguous or even inaccurate categorical information about the visual stimuli.
First, M/EEG signals, images, and text pass through their respective encoders to extract neural features, visual embeddings, and semantic embeddings.
For M/EEG-image alignment, it maps neural features to visual embeddings, preserving spatial-temporal dynamics of brain responses.
For M/EEG-text alignment, we incorporate a dual-text alignment strategy comprising both \textit{Coarse-grained text} and \textit{Fine-grained text} and align the neural features with semantic embeddings using a text projector.
The coarse-grained text contains only category-level labels that provide categorical priors but lacks fine visual details (e.g., color, shape, orientation).
In contrast, fine-grained text, generated by pre-trained large multimodal models (LMMs),  offers rich perceptual descriptions but may contain ambiguous categorical information.
Therefore, coarse-grained and fine-grained text can complement each other, enabling the model to learn more effective neural semantic representations.
% We achieved the multimodal alignment of M/EEG signals with images and text by introducing an image projector and a text projector after the M/EEG encoder.
% After multi-modal alignment, it enables the extraction of M/EEG visual representations and M/EEG semantic representations, which are then utilized for visual stimulus retrieval and classification tasks respectively.

Subsequently, to align the M/EEG signals with images and text, we introduce separate image and text projectors following the M/EEG encoder. These projectors extract neural visual and semantic representations, which are then applied to the visual stimulus retrieval and classification tasks.
For visual stimulus reconstruction, 
% the pre-trained M/EEG encoder is used to extract M/EEG visual representations that are already aligned with image embeddings and text embeddings.
% Then, the M/EEG visual representations and M/EEG semantic representations are subsequently mapped to CLIP image embeddings and prompt embeddings extracted from coarse-grained and fine-grained text separately.
these neural visual and semantic representations are then mapped to CLIP image embeddings and prompt embeddings, derived from the coarse- and fine-grained text, respectively.
Specifically, we employ a diffusion prior~\citep{rameshHierarchicalTextconditionalImage2022} to map the neural visual representations to CLIP image embeddings.
Inspired by the learned queries in BLIP2~\citep{liBlip2BootstrappingLanguageimage2023}, we leverage the Q-Former to map the neural semantic representations to the fused prompt embeddings, derived from both types of textural information.
Finally, these mapped neural embeddings serve as dual conditions for the pre-trained diffusion model, which guides the image reconstruction process from visual and semantic perspectives.

% contributions
The main contributions of this paper are as follows:
\begin{itemize}
    \item
    % We propose a unified multitask framework capable of simultaneously performing zero-shot M/EEG-based visual stimulus retrieval, classification, and reconstruction, where the tasks mutually enhance each other. 
    We present a unified multitask learning framework that synergistically integrates zero-shot M/EEG-based visual stimulus retrieval, classification, and reconstruction through joint optimization.
    % Our integrated framework achieves superior performance compared to training separate models for each task individually.
    This multitask framework outperforms conventional single-task approaches, establishing a new paradigm for neural signal processing with mutually reinforcing feature learning.
    \item
    We propose a novel multimodal alignment strategy that integrates M/EEG, images, and text to simultaneously extract neural visual and semantic representations. The proposed dual-granularity text integration facilitates multimodal alignment, which not only enables the extraction of neural semantic representations but also enhances the extraction of neural visual representations.
    \item
    We separately extract neural visual representations and neural semantic representations and use them as dual conditional inputs to the diffusion model. These representations guide the image generation from both visual and semantic perspectives, ensuring a more comprehensive and accurate reconstruction.
    \item
    We conducted extensive quantitative and qualitative validation experiments on two datasets, covering both EEG and MEG modalities. The experimental results demonstrate the superior performance of our method.
\end{itemize}

The remainder of this paper is structured as follows.
We provide a brief overview of the related works in Section~\ref{sec:RelatedWorks}.
Section~\ref{sec:Method} presents a detailed description of our proposed method.
The experiments and results are illustrated in Section~\ref{sec:ExperimentsResults}.
A careful discussion is presented in Section~\ref{sec:Discussion}.
Finally, we reach a conclusion in Section~\ref{sec:Conclusion}.

	%% 3.Related works
	\section{Related Works}
        \label{sec:RelatedWorks}
While early visual decoding predominantly relied on fMRI, recent advances extend to M/EEG paradigms. Therefore, we review related works on fMRI-based and M/EEG-based visual decoding methods
\subsection{Visual Decoding from fMRI Data}
In earlier studies, researchers employed linear regression to map fMRI data to handcrafted image features or to image features extracted using pre-trained models for visual decoding\citep{kayIdentifyingNaturalImages2008,nishimotoReconstructingVisualExperiences2011,wenNeuralEncodingDecoding2018}.
With the development of generative adversarial nets (GAN)~\citep{goodfellowGenerativeAdversarialNets2014}, some studies~\citep{renReconstructingSeenImage2021,ozcelikReconstructionPerceivedImages2022,linMindReaderReconstructing2022} used the aligned fMRI embeddings as conditional inputs to conditional GAN~\citep{casanovaInstanceconditionedGan2021,karrasAnalyzingImprovingImage2020} for image reconstruction.
% karrasAnalyzingImprovingImage2020
More recently, diffusion models~\citep{rombachHighresolutionImageSynthesis2022} have emerged as highly effective and powerful tools for high-quality image generation.
Numerous studies have leveraged pre-trained diffusion models for fMRI visual stimulus reconstruction, resulting in high-resolution images with exceptional fidelity.
Takagi \textit{et al.}~\citep{takagiHighresolutionImageReconstruction2023} employed Stable Diffusion for visual stimulus reconstruction with fMRI signals from different visual cortex.
Chen \textit{et al.}~\citep{chenSeeingBrainConditional2023} developed a two-stage framework MinD-Vis which first uses mask signal modeling for pre-training to learn fMRI representations and then finetunes a latent diffusion model through double conditioning for image reconstruction.
Ozcelik \textit{et al.}~\citep{ozcelikNaturalSceneReconstruction2023} used ridge regression models to map fMRI signals separately to VAE latent variables, CLIP image embeddings, and CLIP text embeddings and then employed a versatile diffusion model~\citep{xuVersatileDiffusionText2023} for image reconstruction.
Scotti \textit{et al.}~\citep{scottiReconstructingMindsEye2024} proposed a framework called MindEye, which consists of two parallel submodules for visual stimulus retrieval and reconstruction.
Additionally, the model integrates a high-level semantic pipeline and a low-level perceptual pipeline for image generation.
Subsequently, Scotti \textit{et al.}~\citep{scottiMindEye2SharedSubjectModels2024} presented an enhanced model MindEye2 which achieves competitive decoding performance with just one hour of data comparable to using a subject’s full dataset through multi-subject pre-training.

% However, most of these approaches only map fMRI signals to the pre-trained visual-text embeddings or VAE latents, without proposing a multitask framework that can be simultaneously applied to visual stimulus retrieval, classification, and reconstruction tasks after multimodal per-training.

\subsection{Visual Decoding from M/EEG Data}

% previous EEG method and their limitations
Due to the high temporal resolution and convenience of EEG signals, researchers have attempted to use EEG signals for real-time visual decoding.
Spampinato \textit{et al.} proposed an EEG dataset~\citep{spampinatoDeepLearningHuman2017a} for visual object analysis and conducted a series of works~\citep{palazzoGenerativeAdversarialNetworks2017,kavasidisBrain2ImageConvertingBrain2017,tirupatturThoughtVizVisualizingHuman2018,jiangBrainmediaDeepFramework2020}, including EEG visual stimulus decoding and reconstruction using VAE and GAN on this dataset.
However, Li \textit{et al.}~\citep{liPerilsPitfallsBlock2020} pointed out a flaw in the dataset, namely that all stimuli of a given class are presented together.
This results in visual stimulus decoding relying on block-level temporal correlations which are present in all EEG data, rather than the visual information contained in the EEG signals.

% Things-EEG dataset
Recently, with the release of a new large-scale and rigorous Things-EEG dataset~\citep{giffordLargeRichEEG2022}, researchers are now able to perform zero-shot visual stimulus decoding and reconstruction on this dataset.
Song \textit{et al.}~\citep{songDecodingNaturalImages2024} designed an EEG-image contrastive learning framework, NICE, along with two plug-and-play modules that can capture spatial correlations for EEG visual decoding.
NICE++~\citep{song2025recognizing} further incorporated language guidance to capture the semantic information embedded within EEG signals.
Du \textit{et al.}~\citep{duDecodingVisualNeural2023} proposed a multimodal framework BraVL for visual neural representation learning, which utilizes mutual information regularization to align brain signals, images, and text.
However, BraVL is primarily designed for fMRI data and is limited to the visual stimulus classification task.
Wei \textit{et al.}~\citep{weiMB2CMultimodalBidirectional2024} employed contrastive learning and bidirectional cycle consistency to align the feature distributions of EEG and images for visual classification and reconstruction.
% The aligned EEG features were then used as conditional inputs to StyleGAN~\citep{karrasStylebasedGeneratorArchitecture2019} and Stable Diffusion XL (SDXL)~\citep{podell2023sdxlimprovinglatentdiffusion} for image reconstruction.
Based on MEG, Benchetrit\textit{et al.}~\citep{benchetritBrainDecodingRealtime2024} proposed a real-time visual stimulus retrieval and reconstruction framework.
Fu \textit{et al.}~\citep{fuBrainVisExploringBridge2024} developed a novel EEG image reconstruction framework BrainVis which aligns EEG time-frequency embeddings with the interpolation of both coarse-grained and fine-grained text embeddings.
The aligned EEG embeddings serve as conditions of cascaded diffusion models for image reconstruction.
% However, BrainVis does not leverage image modality data for multimodal alignment and it is limited to the visual reconstruction task.
Li \textit{et al.}~\citep{liVisualDecodingReconstruction2024} presented a novel brain decoding framework that can simultaneously perform zero-shot visual stimulus retrieval, classification, and reconstruction tasks based on M/EEG data.
Actually, the work trains a separate M/EEG encoder for each of the three tasks, rather than employing a unified multitask framework.
In conclusion, these M/EEG-based visual decoding approaches are limited to single or dual-task frameworks and do not leverage text for multimodal alignment to simultaneously extract neural visual representations and neural semantic representations.

        %% 4.Methods
	\section{Methods}
        \label{sec:Method}
\subsection{Overall Architecture and Problem Definition}
\begin{figure*}[width=\textwidth,t]
    \centering
    \includegraphics[width=1.0\textwidth]{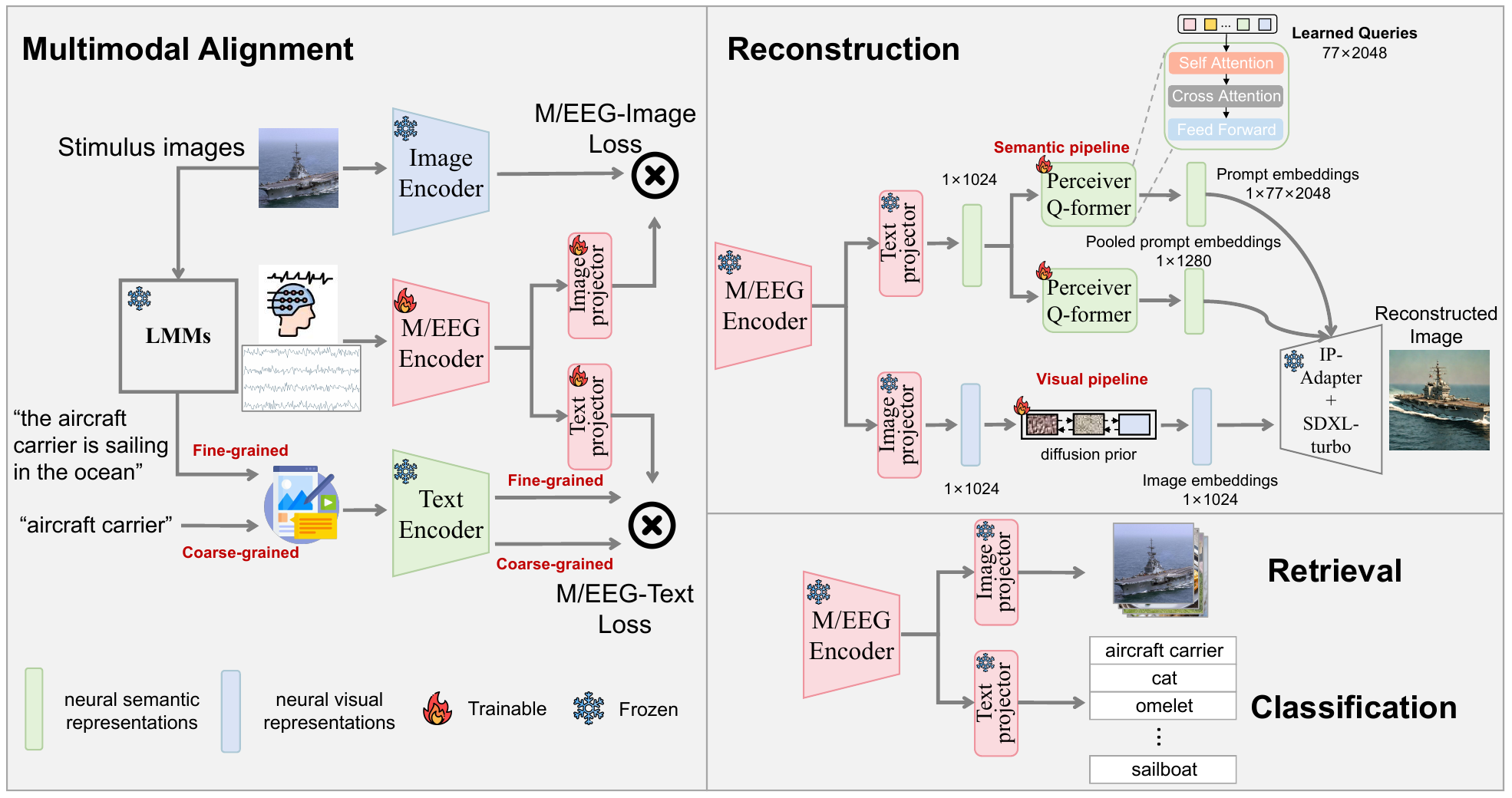}
    \caption{The proposed UMind framework enables zero-shot visual decoding from M/EEG signals, which simultaneously performs visual stimulus retrieval, classification, and reconstruction. It comprises three key components: a multimodal alignment module, a visual stimulus retrieval and classification module, and a dual conditioned diffusion reconstruction module.
    % The overview of our proposed framework UMind.
    % The entire framework is divided into three parts: multi-modal alignment, visual stimulus retrieval and classification, and image reconstruction.
    % In the multimodal alignment phase, an image projector and a text projector are added after the M/EEG Encoder to align M/EEG data with images and text, respectively, thereby simultaneously extracting neural visual representations and neural semantic representations.
    % The extracted neural visual representations and neural semantic representations are utilized for visual stimulus retrieval and classification, as well as for guiding image generation from visual and semantic perspectives, respectively
    }
\label{fig:main_figure}
\end{figure*}

% This work aims to decode and reconstruct images from brain activities recorded via EEG or MEG signals .
% We propose \textbf{UMind}, a multitask framework designed to simultaneously perform zero-shot visual stimulus retrieval, classification and reconstruction.
We propose \textbf{UMind}, a unified multitask framework for zero-shot visual decoding from M/EEG signals, which simultaneously performs visual stimulus retrieval, classification, and reconstruction.
As illustrated in Fig.~\ref{fig:main_figure}, 
% we propose a multitask framework UMind that simultaneously supports zero= visual stimulus retrieval, classification, and reconstruction.
our architecture comprises three main submodules: a multimodal alignment module, a visual stimulus retrieval and classification module, and a dual conditioned diffusion reconstruction module.
% the visual stimulus retrieval and classification module, and the visual-guided and semantic-guided image reconstruction module.
In the multimodal alignment module, it disentangles the neurocognitive representations by aligning neural signals $\mathbf{X}_b$ with visual stimuli $\mathbf{X}_v$ and textual information.
For textual information, we incorporate dual-textual modality data, including coarse-grained texts ($\mathbf{X}_c$) containing only category information and fine-grained descriptions ($\mathbf{X}_t$) of images generated by a pre-trained image-to-text generation model.
This allows our model to extract both neural visual representations and neural semantic representations, which can be used separately for visual stimulus retrieval and classification.
In the reconstruction module, the obtained neural visual representations and neural semantic representations are mapped to CLIP image embeddings and prompt embeddings derived from coarse- and fine-grained text, respectively.
These mapped embeddings guide the generation of realistic and credible images from both visual and semantic perspectives.

We denote the training set $\mathcal{D}_{train}$ as $(\boldsymbol{X}_b, \boldsymbol{X}_v, \boldsymbol{X}_c, \boldsymbol{X}_t, \boldsymbol{Y})=\left\{\left(\boldsymbol{x}_b^i,\boldsymbol{x}_v^i,\boldsymbol{x}_c^i,\boldsymbol{x}_t^i,\boldsymbol{y}^i\right)\right\}_{i=1}^{N_{train}}$, where $N_{train}$ is the sample size, $\boldsymbol{x}_b^i\in\mathbb{R}^{C\times T}$ is the $i^{th}$ EEG or MEG trial in training data with $C$ electrode channels and $T$ time
points, $\boldsymbol{x}_v^i$ denotes the corresponding $i^{th}$ visual stimulus images, $\boldsymbol{x}_c^i$ denotes the coarse-grained texts containing only the category label, $\boldsymbol{x}_t^i$ denotes the detailed descriptions of the image and $\boldsymbol{y}^i$ denotes the corresponding one-hot category label.
Likewise, the test set $\mathcal{D}_{test}$ is defined as $(\boldsymbol{X}_b^{test}, \boldsymbol{X}_v^{test}, \boldsymbol{X}_c^{test}, \boldsymbol{X}_t^{test}, \boldsymbol{Y}^{test})$.
The labels of test data are with no overlap with the labels in the training data, namely, $\boldsymbol{Y}\cap\boldsymbol{Y}^{test}=\emptyset$.
Therefore, the tasks of visual stimulus retrieval, classification, and reconstruction are zero-shot.
The objective is to pre-train a M/EEG encoder using the multimodal training data $\mathcal{D}_{train}$ that includes M/EEG, images, and text, enabling the pre-trained M/EEG encoder to perform zero-shot visual stimulus retrieval, classification, and reconstruction excellently on the test set $\mathcal{D}_{test}$.

\subsection{Multimodal Alignment}
\begin{figure}
    \centering
    \includegraphics[width=1.0\linewidth]{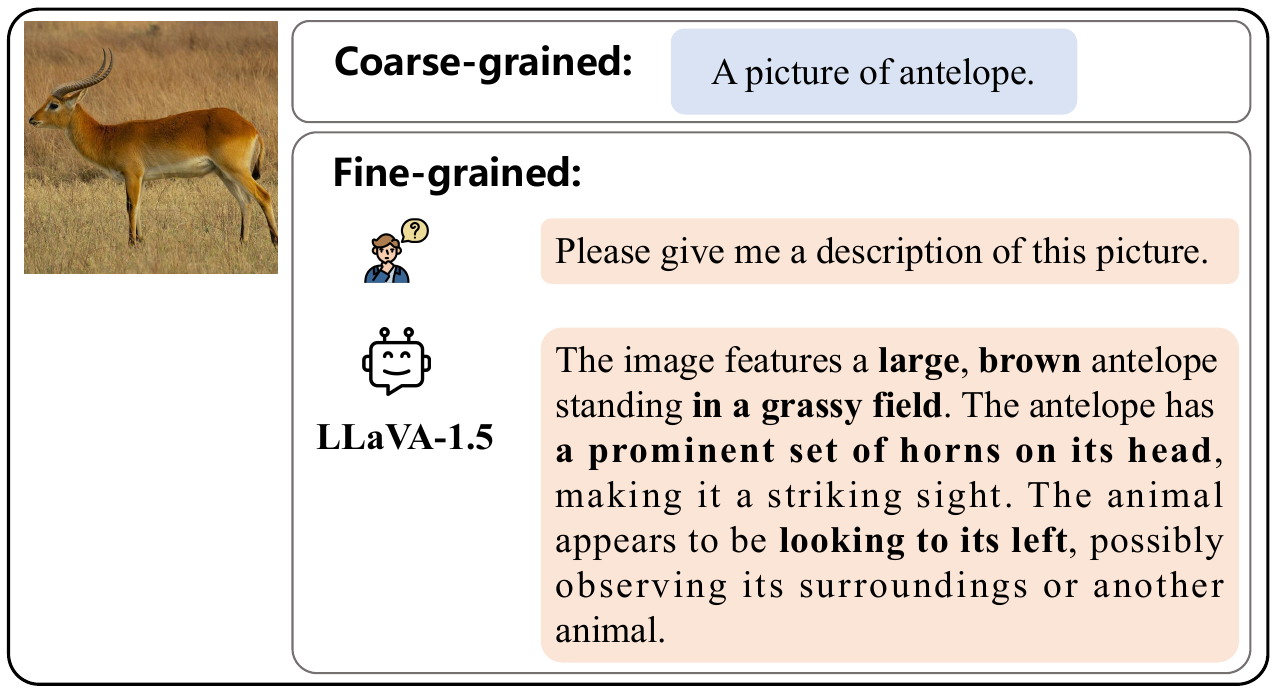}
    \caption{Comparison between fine-grained text generated by LLaVA-1.5 7B and coarse-grained text.}
    \label{fig:coarse_fine_comparison}
\end{figure}
To extract both visual and semantic representations from neural signals (M/EEG), we employ a multimodal framework that bridges neural signals with paired images and text.
% to align the neural signals with the corresponding images and texts.
This framework integrates a M/EEG encoder $f_b(\cdot)$ with two parallel projectors: an image projector $p_v(\cdot)$ and a text projector $p_t(\cdot)$. 
The neural signals are first encoded via $f_b(\cdot)$, then mapped through $p_v(\cdot)$ to derive neural-visual representations $\hat{\boldsymbol{z}}_v^i$ ($\hat{\boldsymbol{z}}_v^i=p_v(f_b(\boldsymbol{x}_b^i))$), which align with the CLIP image embeddings $\boldsymbol{z}_v^i$ from a frozen pre-trained image encoder. 
Simultaneously, the neural-semantic representations generated by $\hat{\boldsymbol{z}}_s^i$ ($\hat{\boldsymbol{z}}_s^i=p_t(f_b(\boldsymbol{x}_b^i))$) are aligned with the CLIP text embeddings extracted by a frozen pre-trained text encoder.
% the image projector and a text projector following the M/EEG encoder.
% The EEG signals are processed through the EEG encoder and the image projector to extract EEG visual representations, which are then aligned with the CLIP image embeddings extracted by the pre-trained image encoder.
% Similarly, the EEG signals are processed through the EEG encoder and the text projector to extract EEG semantic representations, which are aligned with the CLIP text embeddings extracted by the pre-trained text encoder.

In aligning M/EEG with textual information, we employ dual-grained supervision: coarse-grained labels and fine-grained captions, as illustrated in Fig.~\ref{fig:coarse_fine_comparison}.
% As illustrated in Fig.~\ref{fig:coarse_fine_comparison}, the coarse-grained text includes only category information, while the fine-grained text provides additional details such as location, color, and surrounding environment in addition to category information.
% We use LLaVA-1.5~\citep{liuImprovedBaselinesVisual2024} or BLIP-2~\citep{liBlip2BootstrappingLanguageimage2023} to generate fine-grained text descriptions for the images.
We apply LLaVA-1.5~\citep{liuImprovedBaselinesVisual2024} 
% or BLIP-2~\citep{liBlip2BootstrappingLanguageimage2023} 
to generate fine-grained captions (enriched with spatial, chromatic, and contextual details).
By combining fine-grained and coarse-grained text, we can extract rich semantic information with more detailed descriptions, while also enhancing the extraction of visual representations.
We denote the coarse- and fine-grained image embeddings as $\boldsymbol{z}_c^i$ and $\boldsymbol{z}_t^i$, respectively. 
% We denote the EEG encoder as $f_b(\cdot)$. 
% The image projector and text projector defined as $p_v(\cdot)$ and $p_t(\cdot)$, are connected to the EEG encoder respectively.
% The pre-trained CLIP image encoder and CLIP text encoder are defined as $f_v(\cdot)$ and $f_t(\cdot)$, respectively.
% The EEG visual representations $\hat{\boldsymbol{z}}_v^i=p_v(f_b(\boldsymbol{x}_b^i))$ are aligned with the image embeddings $\boldsymbol{z}_v^i=f_v(\boldsymbol{x}_v^i)$ and EEG semantic representations $\hat{\boldsymbol{z}}_s^i=p_t(f_b(\boldsymbol{x}_b^i))$ are aligned with text embeddings $\boldsymbol{z}_c^i=f_t(\boldsymbol{x}_c^i),\boldsymbol{z}_t^i=f_t(\boldsymbol{x}_t^i)$ through contrastive learning:
The neural-visual and neural-semantic representations are aligned with their multimodal counterparts using contrastive learning loss with  
\begin{equation}
    \label{eq:contrast_v}
    \mathcal{L}_{CLIP_V}=-\frac{1}{B}\sum_{i=1}^B\log\frac{\exp(s(\hat{\boldsymbol{z}}_v^i,\boldsymbol{z}_v^i)/\tau)}{\sum_{j=1}^B\exp(s(\hat{\boldsymbol{z}}_v^i,\boldsymbol{z}_v^j)/\tau)}
\end{equation}

\begin{equation}
    \label{eq:contrast_t1}
    \mathcal{L}_{CLIP_{T1}}=-\frac{1}{B}\sum_{i=1}^B\log\frac{\exp(s(\hat{\boldsymbol{z}}_s^i,\boldsymbol{z}_c^i)/\tau)}{\sum_{j=1}^B\exp(s(\hat{\boldsymbol{z}}_s^i,\boldsymbol{z}_c^j)/\tau)}
\end{equation}

\begin{equation}
    \label{eq:contrast_t2}
    \mathcal{L}_{CLIP_{T2}}=-\frac{1}{B}\sum_{i=1}^B\log\frac{\exp(s(\hat{\boldsymbol{z}}_s^i,\boldsymbol{z}_t^i)/\tau)}{\sum_{j=1}^B\exp(s(\hat{\boldsymbol{z}}_s^i,\boldsymbol{z}_t^j)/\tau)}
\end{equation}

\begin{equation}
    \label{eq:contrast_t}
    \mathcal{L}_{CLIP_T} = \left(\mathcal{L}_{CLIP_{T1}}+\mathcal{L}_{CLIP_{T2}}\right)/2
\end{equation}

% where $\mathcal{L}_{CLIP_V}$ is the contrastive loss between EEG and images, $\mathcal{L}_{CLIP_T}$ is the contrastive loss between EEG and text, 
where $\mathcal{L}_{CLIP_V}$ and $\mathcal{L}_{CLIP_T}$ are the contrastive loss between neural representations with image and text embeddings, respectively. $s$ denotes cosine similarity, $\tau$ is the temperature parameter and $B$ is the batch size.

Additionally, we also compute the Mean Squared Error (MSE) loss between the neural representations and the image and text embeddings: 
% for the subsequent image reconstruction task:
\begin{equation}
    \label{eq:mse_v}
    \mathcal{L}_{MSE_V}=\frac{1}{Bd}\sum_{i=1}^B\lVert \boldsymbol{z}_v^i-\hat{\boldsymbol{z}}_v^i\rVert_2^2
\end{equation}

\begin{equation}
    \label{eq:mse_t}
    \mathcal{L}_{MSE_T}=\frac{1}{Bd}\sum_{i=1}^B\left(\lVert \boldsymbol{z}_c^i-\hat{\boldsymbol{z}}_s^i\rVert_2^2+\lVert \boldsymbol{z}_t^i-\hat{\boldsymbol{z}}_s^i\rVert_2^2\right)/2
\end{equation}
% where $\mathcal{L}_{MSE_V}$ is the MSE loss between neural visual representations and image embeddings, $\mathcal{L}_{MSE_T}$ is the MSE loss between neural semantic representations and text embeddings, 
where $d$ is the dimension of representations.

During training, both the CLIP image and text encoder are frozen.
We optimize the M/EEG encoder, image projector, and text projector by combining the contrastive loss and MSE loss.
The overall loss function is: 
\begin{equation}
    \label{eq:overall_loss}
    \mathcal{L}_{all}=\alpha(\mathcal{L}_{CLIP_V}+\beta\mathcal{L}_{MSE_V})+(1-\alpha)(\mathcal{L}_{CLIP_T}+\beta\mathcal{L}_{MSE_T})
\end{equation}
where $\alpha$ and $\beta$ are the hyperparameters that control the balance between the M/EEG-image alignment loss and the M/EEG-text alignment loss, and between the contrastive loss and the MSE loss, respectively.

\subsection{Visual Stimulus Retrieval and Classification}
In the test phase, after multimodal alignment,  we employ cosine similarity to match the extracted neural-visual and semantic representations with image and category templates, facilitating zero-shot visual stimulus retrieval and classification. 
These templates are constructed using images and coarse-level label texts that do not overlap with the training samples. 
% The images and coarse-level label texts with no overlap with the training samples are utilized to construct the templates.
Each image and its corresponding text are processed separately through pre-trained CLIP image and text encoders to build the templates.

\subsection{Dual Guidance for Image Reconstruction}
Through the multimodal alignment pre-training, we can simultaneously extract neural-visual representations and neural-semantic representations, which jointly guide visual stimulus reconstruction. 
% using the frozen M/EEG encoder, image projector, and text projector.
Inspired by~\citep{liVisualDecodingReconstruction2024}, we integrate the pre-trained IP-Adapter~\citep{ye2023ipadaptertextcompatibleimage} with the SDXL-Turbo~\citep{sauerAdversarialDiffusionDistillation2025} to enable dual-modality conditioned image generation.
SDXL-Turbo utilizes Adversarial Diffusion Distillation (ADD) to generate high-quality images using only 1–4 sampling steps.
% With the addition of the IP-Adapter, SDXL-Turbo not only enables text-to-image generation but also achieves image prompt capability for image generation.
The addition of the IP-Adapter enables both text-to-image generation and image prompt capabilities, allowing for guiding image reconstruction from both visual and semantic perspectives using the extracted neural representations.

% \subsubsection{Visual guidance for reconstruction}
\textbf{Visual guidance:}
% Inspired by~\citep{rameshHierarchicalTextconditionalImage2022,scottiReconstructingMindsEye2024}, contrastive learning solely encourages the neural representations to align with the direction of the corresponding CLIP image embeddings in the vector space.
While contrastive learning aligns neural-visual representations with CLIP image embeddings, it still lacks explicit spatial reconstruction~\citep{scottiReconstructingMindsEye2024}. 
Therefore, we train a diffusion prior to map the neural visual representations into the CLIP image space.
This prior enables the pre-trained diffusion model to perform image reconstruction from the visual perspective.

% \subsubsection{Semantic guidance for reconstruction}
\textbf{Semantic guidance:}
% Due to the low dimensionality of neural representations, directly decoding captions from neural representations aligned with the CLIP image space to guide image reconstruction is unstable~\citep{liVisualDecodingReconstruction2024}.
Direct caption decoding from low-dimensional neural-semantic representations proves unstable due to information loss in alignment~\citep{liVisualDecodingReconstruction2024}.
% Therefore, we attempt to map the extracted neural semantic representations to the CLIP text space to directly guide image generation, thus avoiding the instability associated with using an image-to-text model to generate captions from neural representations.
Inspired by BLIP-2~\citep{liBlip2BootstrappingLanguageimage2023}, we project these neural-semantic representations to the CLIP text space using a Q-Former block with a set of learnable query vectors.
Specifically, the learnable query vectors $\boldsymbol{z}_q$ first pass through a self-attention layer~\citep{vaswaniAttentionAllYou2017}:
% the learnable query vectors $\boldsymbol{z}_q$ iteratively attend to neural-semantic features $\hat{\boldsymbol{z}}_s^i$  
% SDXL-Turbo is a distilled version of SDXL 1.0~\citep{podell2023sdxlimprovinglatentdiffusion}, which employs CLIP ViT-L and OpenCLIP ViT-bigG as pre-trained text encoders for text conditioning.
% We utilize the two pre-trained text encoders of SDXL-Turbo to extract features for coarse-grained and fine-grained text, where we concatenate the penultimate text encoder outputs along the channel axis, resulting in prompt embeddings with a dimensionality of $\mathbb{R}^{77\times2048}$.
% Inspired by BLIP-2~\citep{liBlip2BootstrappingLanguageimage2023}, we use the image block in Q-Former to map the neural semantic representations with a dimensionality of $\mathbb{R}^{1024}$ to prompt embeddings with a dimensionality of $\mathbb{R}^{77\times2048}$, using a set of learnable query vectors.
% The number of learnable query vectors is set to the number of tokens in the prompt embeddings, for example, 77.

% In our Q-Former block, we denote the learnable query embeddings as $\boldsymbol{z}_q$.
% First, the learnable query embeddings pass through a self-attention layer~\citep{vaswaniAttentionAllYou2017}:
\begin{equation}
    \label{eq:Q_selfattention}
    \boldsymbol{z}_q^{\prime}=\mathrm{SelfAttention}(\boldsymbol{z}_q)
\end{equation}

Next, the learnable query embeddings $\boldsymbol{z}_q^{\prime}$ attend to neural-semantic features $\hat{\boldsymbol{z}}_s^i$:
% and neural semantic representations $\boldsymbol{z}_s$ pass through a cross-attention layer, 
% which enforces the learnable query embeddings to learn the semantic representations and 
% expanding the dimensions from $\mathbb{R}^{1024}$ to $\mathbb{R}^{77\times2048}$:
\begin{equation}
    \label{eq:QKV_crossattention}
    Q=W_Q\cdot\boldsymbol{z}_q^{\prime},K=W_K\cdot \hat{\boldsymbol{z}}_s,V=W_V\cdot\hat{\boldsymbol{z}}_s
\end{equation}

\begin{equation}
    \label{eq:out_crossattention}
    \boldsymbol{z}_q^{\prime\prime}=\mathrm{Attention}(Q,K,V)=\mathrm{softmax}\left(\frac{QK^T}{\sqrt{d}}\right)\cdot V
\end{equation}
where $W_Q,W_K$ and $W_V$ are learnable projection matrices.
% , with $W_K$ and $W_V$ mapping $\hat{\boldsymbol{z}}_s$ from $\mathbb{R}^{1024}$ to $\mathbb{R}^{2048}$.

Finally, the learnable query embeddings $\boldsymbol{z}_q^{\prime\prime}$ pass through a Feed-Forward Network (FFN) to produce the outputs.
\begin{equation}
    \label{eq:FFN}
    \boldsymbol{z}_q^{prompt}=\mathrm{FFN}(\boldsymbol{z}_q^{\prime\prime})
\end{equation}

Additionally, SDXL-Turbo conditions the model on the pooled prompt embedding.
Similarly, we utilize the Q-Former to map $\hat{\boldsymbol{z}}_s$ into pooled prompt embeddings, denoted as $\boldsymbol{z}_q^{pool\_prompt}$, which guide image reconstruction from the semantic perspective with the prompt embeddings together.
In this way, the proposed dual-guidance approach enhances image reconstruction by leveraging both visual and semantic information.
        
        %% 5.Experiments
	\section{Experiments}
        \label{sec:ExperimentsResults}
\begin{table*}[width=\textwidth, htbp]
    \centering
    % \captionsetup{font=small}
    \caption{Retrieval and classification accuracy (\%) of different methods on THINGS-EEG dataset}
    \setlength{\tabcolsep}{1.5pt}
    % \small
    % \footnotesize
    \scriptsize
    % \tiny
    \begin{tabular}{lcccccccccccccccccccccc}
        \toprule
        \multirow{2}*{Methods} & \multicolumn{2}{c}{Subject 1} & \multicolumn{2}{c}{Subject 2} & \multicolumn{2}{c}{Subject 3} & \multicolumn{2}{c}{Subject 4} & \multicolumn{2}{c}{Subject 5} & \multicolumn{2}{c}{Subject 6} & \multicolumn{2}{c}{Subject 7} & \multicolumn{2}{c}{Subject 8} & \multicolumn{2}{c}{Subject 9} & \multicolumn{2}{c}{Subject 10} & \multicolumn{2}{c}{Average} \\
        \cmidrule(l){2-3} \cmidrule(l){4-5} \cmidrule(l){6-7} \cmidrule(l){8-9} \cmidrule(l){10-11} \cmidrule(l){12-13} \cmidrule(l){14-15} \cmidrule(l){16-17} \cmidrule(l){18-19} \cmidrule(l){20-21} \cmidrule(l){22-23}
        & top-1 & top-5 & top-1 & top-5 & top-1 & top-5 & top-1 & top-5 & top-1 & top-5 & top-1 & top-5 & top-1 & top-5 & top-1 & top-5 & top-1 & top-5 & top-1 & top-5 & top-1 & top-5  \\
        \midrule
        \multicolumn{23}{c}{Retrieval accuracy} \\
        \midrule
        NICE & 21.50 & 53.50 & 21.00 & 51.00 & 27.50 & 62.50 & 30.00 & 63.50 & 13.00 & 36.00 & 22.50 & 56.00 & 25.50 & \textbf{59.50} & 38.50 & 68.50 & 21.00 & 57.00 & 23.00 & 64.50 & 24.35 & 57.20 \\
        ATM & 20.50 & \textbf{58.00} & 18.00 & 47.50 & 25.00 & 60.00 & 27.50 & 58.00 & 15.50 & 42.00 & 27.50 & 63.50 & 24.00 & 53.00 & 41.00 & 72.00 & 21.50 & 51.00 & 36.50 & 69.50 & 25.70 & 57.45 \\
        MB2C & 23.67 & 56.33 & 22.67 & 50.50 & 26.33 & 60.17 & 34.83 & 67.00 & 21.33 & \textbf{53.00} & 31.00 & 62.33 & 25.00 & 54.83 & 39.00 & 69.33 & 27.50 & 59.33 & 33.17 & 70.83 & 28.45 & 60.37  \\
        
        UMind & \textbf{27.00} & 56.00 & \textbf{32.00} & \textbf{70.00} & \textbf{34.00} & \textbf{70.00} & \textbf{36.00} & \textbf{70.50} & \textbf{23.00} &      50.50     & \textbf{32.50} & \textbf{68.50} & \textbf{28.00} & 59.00 & \textbf{46.50} & \textbf{80.00} & \textbf{37.50} & \textbf{66.50} & \textbf{42.00} & \textbf{76.00} & \textbf{33.85} & \textbf{66.70} \\
        \midrule
        \multicolumn{23}{c}{Classification accuracy} \\
        \midrule
        BraVL & 6.11 & 17.89 & 4.90 & 14.87 & 5.58 & 17.38 & 4.96 & 15.11 & 4.01 & 13.39 & 6.01 & 18.18 & 6.51 & 20.35 & 8.79 & 23.68 & 4.34 & 13.98 & 7.04 & 19.71 & 5.82 & 17.45 \\
        
        NICE & \textbf{9.00} & 27.50 & \textbf{10.50} & 24.50 & 10.00 & 37.50 & 11.50 & 35.50 & 6.50 & 20.50 & 9.00 & 28.50 & 9.50 & \textbf{31.00} & 11.50 & 37.50 & 7.00 & 29.00 & 11.50 & 31.50 & 9.60 & 30.30 \\
        
        ATM & 4.50 & 15.50 & 1.50 & 7.00 & 5.00 & 17.00 & 4.50 & 15.50 & 2.50 & 11.00 & 4.50 & 12.50 & 6.50 & 18.00 & 8.00 & 23.00 & 2.50 & 10.00 & 5.50 & 18.50 & 4.50 & 14.80 \\
        
        UMind & 7.50 & \textbf{31.00} & 7.50 & \textbf{32.00} & \textbf{15.50} & \textbf{37.50} & \textbf{17.00} & \textbf{38.50} & \textbf{8.00} & \textbf{22.50} & \textbf{10.50} & \textbf{37.50} & \textbf{11.50} & 30.00 & \textbf{16.00} & \textbf{46.00} & \textbf{10.00} & \textbf{32.50} & \textbf{16.50} & \textbf{36.50} & \textbf{12.00} & \textbf{34.40} \\
        \bottomrule
    \end{tabular}
    \label{tab:retrieval_cls_comparison_eeg}
\end{table*}
%%%%%%%%%%%%%%%%%%%%%%%%%%%%%%%%%%%%%%%%%%%%%%
\begin{table*}[htbp]
    \centering
    % \captionsetup{font=small}
    \caption{Retrieval and classification accuracy (\%) of different methods on THINGS-MEG dataset}
    \setlength{\tabcolsep}{8pt}
    % \small
    % \footnotesize
    % \scriptsize
    % \tiny
    \begin{tabular}{lcccccccccc}
        \toprule
        \multirow{2}*{Methods} & \multicolumn{2}{c}{Subject 1} & \multicolumn{2}{c}{Subject 2} & \multicolumn{2}{c}{Subject 3} & \multicolumn{2}{c}{Subject 4} & \multicolumn{2}{c}{Average} \\
        \cmidrule(l){2-3} \cmidrule(l){4-5} \cmidrule(l){6-7} \cmidrule(l){8-9} \cmidrule(l){10-11}
        & top-1 & top-5 & top-1 & top-5 & top-1 & top-5 & top-1 & top-5 & top-1 & top-5 \\
        \midrule
        \multicolumn{11}{c}{Retrieval accuracy} \\
        \midrule
        NICE & 12.00 & 36.50 & 23.50 & 61.00 & 18.00 & 50.00 & 14.00 & 38.00 & 16.87 & \textbf{46.38} \\
        
        ATM & 11.50 & 32.00 & 29.00 & 65.50 & 24.00 & 48.50 & 9.00 & 30.50 & 18.38 & 44.13  \\
        
        UMind & 7.50 & 30.00 & 38.50 & 69.50 & 23.00 & 49.50 & 8.50 & 30.00 & \textbf{19.38} & 44.75  \\
        \midrule
        \multicolumn{11}{c}{Classification accuracy} \\
        \midrule
        NICE & 4.00 & 16.50 & 9.00 & 33.00 & 10.50 & 26.50 & 5.50 & 17.00 & 7.25 & 23.25 \\
        
        ATM & 1.00 & 13.00 & 10.50 & 28.50 & 7.50 & 21.00 & 2.50 & 10.00 & 5.38 & 18.13  \\
        
        UMind & 5.50 & 18.50 & 11.00 & 34.50 & 13.00 & 32.50 & 6.50 & 22.50 &\textbf{9.00} & \textbf{27.00}  \\
        \bottomrule
    \end{tabular}
    \label{tab:retrieval_cls_comparison_meg}
\end{table*}

\subsection{Datasets and Preprocessing}
\textbf{THINGS-EEG} The THINGS-EEG dataset~\citep{giffordLargeRichEEG2022} comprises EEG recordings from ten participants acquired via 63 electrodes at 1000 Hz during a rapid serial visual presentation (RSVP) task.
Each of the 10 participants completed four identical experimental sessions, resulting in a total of 10 datasets.
For each dataset, the training set consists of 1654 categories, with 10 images per category, repeated 4 times.
The test set includes 200 categories, with one image per category, repeated 80 times.
Each dataset contains a total of 82160 image trials.

\textbf{THINGS-MEG} The THINGS-MEG dataset~\citep{hebartTHINGSdataMultimodalCollection2023} consists of MEG data from 4 subjects with 12 MEG sessions, recorded using 271 channels at a 1200 Hz sampling rate.
The image stimuli were presented for 500ms, followed by a variable fixation period of 1000$\pm$200ms.
The training dataset has 1854 concepts $\times$ 12 images $\times$ 1 repetition, and the
test dataset has 200 concepts $\times$ 1 image $\times$ 12 repetitions.
To create the zero-shot task, we removed 200 test concepts from the training set.
The MEG data were segmented into trials, spanning from 0 to 1000 ms after stimulus onset.

For preprocessing, the EEG is filtered between 0.1 Hz and 100 Hz. EEG data were sampled from 0 to 1000 ms after stimulus onset, baseline corrected, and downsampled to 250 Hz, followed by multivariate noise normalization~\citep{guggenmosMultivariatePatternAnalysis2018} on the training data. Similarly, MEG signals were filtered between 0.1 Hz and 100 Hz, downsampled from 1200 Hz to 200 Hz. To improve the signal-to-noise ratio, EEG and MEG trials were averaged across repetitions for each image.

\subsection{Experiments Settings}
% Our method was implemented in Python 3.10 using the PyTorch framework, and all experiments were conducted on a single NVIDIA GeForce RTX 4090 GPU.
Our method was implemented in Python 3.10 using the PyTorch framework, with all experiments conducted on an NVIDIA GeForce RTX 4090 GPU.
% Following~\citep{songDecodingNaturalImages2024}, we randomly selected 740 samples from the training set as a validation set to identify the optimal model during training.
% Testing was performed only once after the completion of training.
Following~\citep{songDecodingNaturalImages2024}, 
$740$ samples from the training set were randomly selected as a validation set for model optimization.
% The training process utilized the Adam optimizer, setting the learning rate, $\beta_1$, and $\beta_2$ at $2\times10^{-4}$, $0.5$, and $0.999$, respectively.、
The model was trained using the Adam optimizer with a learning rate of $2\times10^{-4}$ for 100 epochs with a batch size of $256$.
% The model was trained with a batch size of 256 for a total of 100 epochs.
In the contrastive loss, the temperature parameter $\tau$ is a learnable parameter and is initially set to 0.07.
The hyperparameters $\alpha$ and $\beta$ in Eq.~(\ref{eq:overall_loss}) were set to $0.5$ and $2$, respectively.
% The model was trained with a batch size of 256 for a total of 100 epochs.
% In the contrastive loss, the temperature parameter $\tau$ was learnable and initialized to $0.07$.

%%%%%%%%%%%%%%%%%%%%%%%%%%%%%%%%%%%%%%%%%%%%%%%%%%%
\begin{figure}
    \centering
    \includegraphics[width=1.0\linewidth]{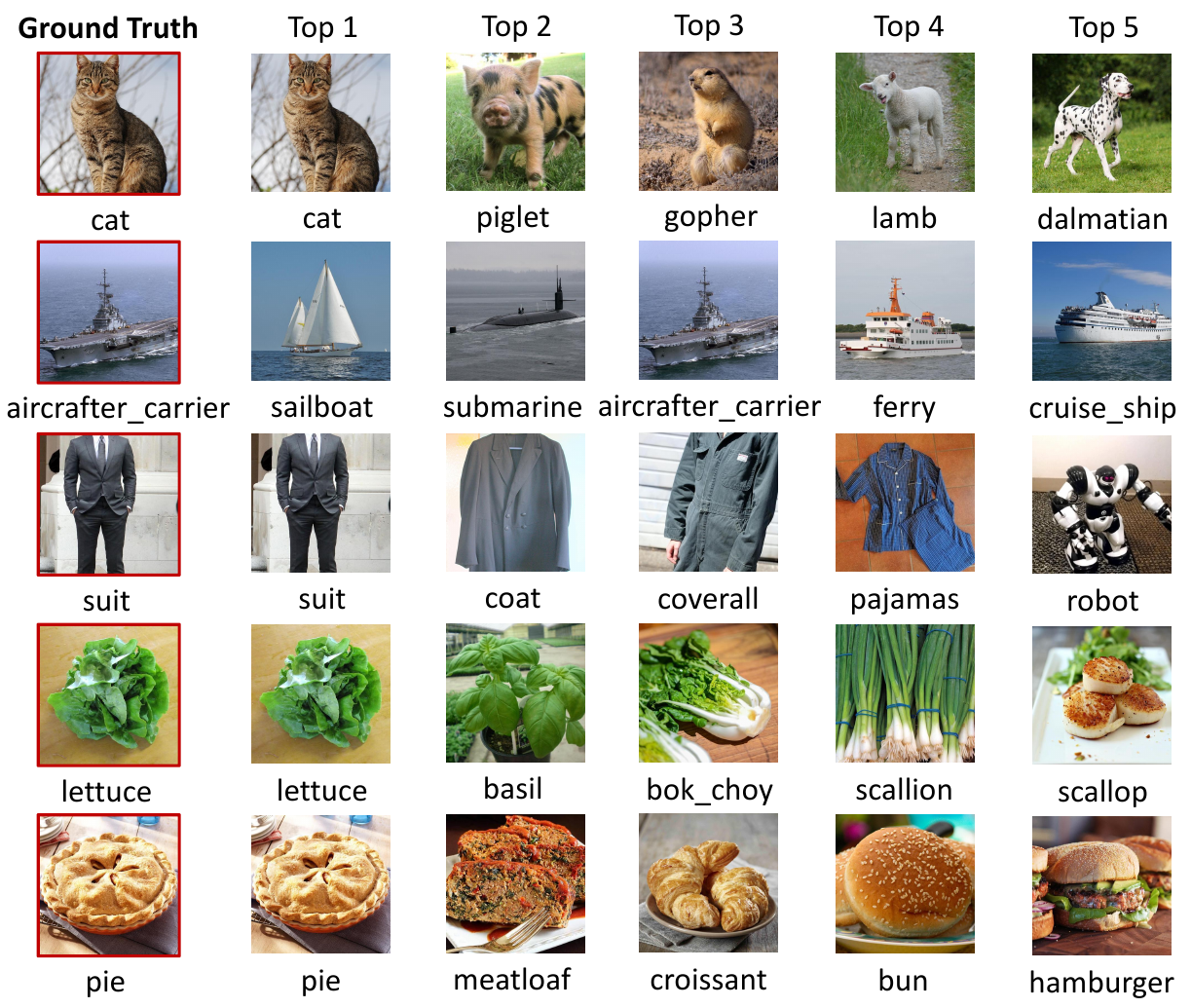}
    \caption{Visualization of top-5 retrieval examples for the retrieval task.}
    \label{fig:retrieval_example}
\end{figure}
%%%%%%%%%%%%%%%%%%%%%%%%%%%%%%%%%%%%%%%%%%%%%%%%%%%
\subsection{Retrieval and Classification Performance}
\subsubsection{Quantitative Comparison Results}
% After completing the multimodal alignment training, we use the pre-trained model to perform zero-shot visual stimuli retrieval and classification on dataset THINGS-EEG.
% To evaluate our method, we compared it 
We evaluated our UMind model through zero-shot visual stimulus retrieval and classification tasks on the THINGS-EEG dataset, comparing it with several state-of-the-art approaches, including BraVL~\citep{duDecodingVisualNeural2023}, NICE~\citep{songDecodingNaturalImages2024}, MB2C~\citep{weiMB2CMultimodalBidirectional2024}, and ATM~\citep{liVisualDecodingReconstruction2024}.
Wilcoxon Signed-Rank Test was employed to evaluate statistical significance.
% Since Li \textit{et al.}~\citep{liVisualDecodingReconstruction2024} did not provide the visual stimulus retrieval and classification results for each subject in their paper, we reproduced their method and tested it only once after the model training was completed to ensure a fair comparison.
As shown in Table~\ref{tab:retrieval_cls_comparison_eeg}, our UMind achieves superior performance for both retrieval and classification tasks.
In the retrieval task, our method achieves a top-1 accuracy of 33.85\% and a top-5 accuracy of 66.70\%, significantly surpassing the chance levels of 0.5\% and 2.5\%, respectively.
In the classification task, our method achieves a top-1 accuracy of 12.00\% and an average top-5 accuracy of 34.40\%, outperforming the existing best method by 2.40\%  (\textit{p}\ \textless\ 0.05) and 4.10\%  (\textit{p}\ \textless\ 0.05), respectively.
These observations highlight the effectiveness of integrating text modality for both retrieval and classification.
% The experimental results demonstrate that introducing the text modality for multimodal alignment not only enables our model to be utilized for both retrieval and classification tasks but also significantly enhances its ability to learn powerful neural visual representations and semantic representations.

To validate generalizability, we tested UMind on the THING-MEG dataset, as shown in Table~\ref{tab:retrieval_cls_comparison_meg}. 
% we conducted experiments on another MEG dataset, THING-MEG.
% The retrieval and classification results are shown in Table~\ref{tab:retrieval_cls_comparison_meg}. 
Our method achieved a top-1 accuracy of 19.38\% for the retrieval task and 9.00\% for the classification task, respectively.
Similarly, our unified multitask framework outperforms individually trained models for each task on the THINGS-MEG dataset, demonstrating the generalizability of our approach.

\subsubsection{Quanlitative Comparison Results}
We randomly showcase several top-5 retrieval results from the retrieval task, as illustrated in Fig.~\ref{fig:retrieval_example}.
% It can be observed that the top-5 results are semantically similar.
These results revealed consistent semantic patterns, 
for example, the retrieval results for ``aircraft carrier" are all related to ships, while those for ``lettuce" are all associated with vegetable-related items.
This indicates that UMind effectively captures semantic details from dual-grained text and neural signals.
% This indicates that our framework can learn the detailed semantic information embedded within the M/EEG signals after integrating coarse-grained and fine-grained text.

%%%%%%%%%%%%%%%%%%%%%%%%%%%%%%%%%%%%%%%%%%%%%%%%%%%
\begin{figure*}[width=\textwidth, t]
    \centering
    \includegraphics[width=1.0\linewidth]{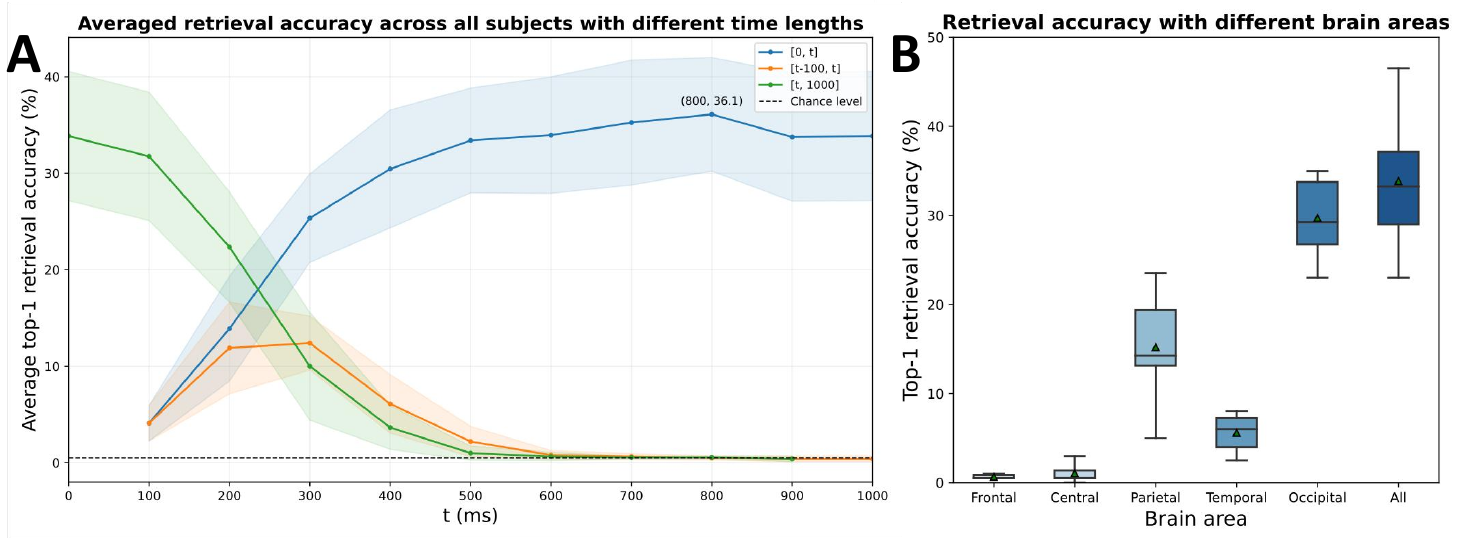}
    \caption{The results of temporal and spatial analysis on THINGS-EEG dataset. (A) The average top-1 retrieval accuracy of all subjects using different EEG time windows: [0, $t$], [$t$-100, $t$], and [$t$, 1000]. (B) Retrieval performance using electrode channels from different brain areas.}
\label{fig:temporal_spatial_analysis}
\end{figure*}
%%%%%%%%%%%%%%%%%%%%%%%%%%%%%%%%%%%%%%%%%%%%%%%%%%%
\subsubsection{Temporal and Spatial Analysis}
To investigate the temporal and spatial characteristics of visual stimulus retrieval, we conducted experiments using different EEG time windows and electrode channels from various brain regions.
% The temporal and spatial analysis results for the retrieval task are presented in Fig.~\ref{fig:temporal_spatial_analysis}, which show similar temporal and spatial characteristics for the classification task.
As shown in Fig.~\ref{fig:temporal_spatial_analysis}(A), three types of time windows were employed: expanding time windows [$0$,$t$], decreasing time windows [$t$,$1000$], and the sliding time windows [$t-100$,$t$], respectively.
It can be observed that the results of the expanding window stabilize at approximately $t=500ms$.
For the sliding window, results exceeding the chance level are primarily observed between $100-500 ms$.
Therefore, the effective information for visual stimulus retrieval is predominantly concentrated within the $0-500 ms$.

As shown in Fig.~\ref{fig:temporal_spatial_analysis}(B), the frontal and central electrodes contribute minimally to visual stimulus retrieval.
In contrast, the occipital, parietal, and temporal electrodes contain varying degrees of effective visual information, ranked from high to low.
When using only occipital electrodes, the average top-1 retrieval accuracy is 19.7\%, with a decrease of 4.15\% (\textit{p}\ \textgreater\ 0.05) compared to using electrodes from all brain regions.
Similar phenomena occur for the classification task.

%%%%%%%%%%%%%%%%%%%%%%%%%%%%%%%%%%%%%%%%%%%%%%%%%
\begin{table*}[width=0.8\textwidth, htbp]
    \centering
    % \captionsetup{font=small}
    \caption{The quantitative comparison of reconstruction performance between our model and other methods.}
    \setlength{\tabcolsep}{3pt}
    % \small
    % \footnotesize
    % \scriptsize
    % \tiny
    \begin{tabular}{l|ccccccccc}
        \toprule
        \multirow{2}*{Datasets} & \multirow{2}*{Methods} & \multicolumn{4}{c}{Low-Level} & \multicolumn{4}{c}{High-Level} \\
        \cmidrule(lr){3-6} \cmidrule(l){7-10} 
         & & PixCorr $\uparrow$ & SSIM $\uparrow$ & AlexNet(2) $\uparrow$ & AlexNet(5) $\uparrow$ & Inception $\uparrow$ & CLIP $\uparrow$  & EfficientNet $\downarrow$ & SwAV $\downarrow$ \\
        \midrule
        \multirow{3}*{THINGS-EEG}
        & MB2C & \textbf{0.188} & 0.333 & $-$ & $-$ & $-$ & $-$ & $-$ & $-$ \\
        
        & ATM & 0.164 & \textbf{0.513} & 0.583 & 0.603 & 0.558 & 0.577 & 0.962 & 0.630 \\
        
        & UMind & 0.156 & 0.390 & \textbf{0.725} & \textbf{0.839} & \textbf{0.744} & \textbf{0.798} & \textbf{0.879} & \textbf{0.560} \\
        \midrule
        \multirow{3}*{THINGS-MEG}
        & B.D. & 0.076 & 0.336 & \textbf{0.736} & \textbf{0.826}  & 0.671 & \textbf{0.767} & $-$ & \textbf{0.584} \\
        
        & ATM & 0.104 & 0.340 & 0.613 & 0.672 & 0.619 & 0.603 & $-$ & 0.651 \\
        
        & UMind & \textbf{0.107} & \textbf{0.360} & 0.700 & 0.808 & \textbf{0.679} & 0.754 & \textbf{0.915} & 0.601 \\
        \bottomrule
    \end{tabular}
    \label{tab:generation_comparison}
\end{table*}
%%%%%%%%%%%%%%%%%%%%%%%%%%%%%%%%%%%%%%%%%%%%%%%%
\begin{figure*}[width=\textwidth, t]
    \centering
    \includegraphics[width=1.0\linewidth]{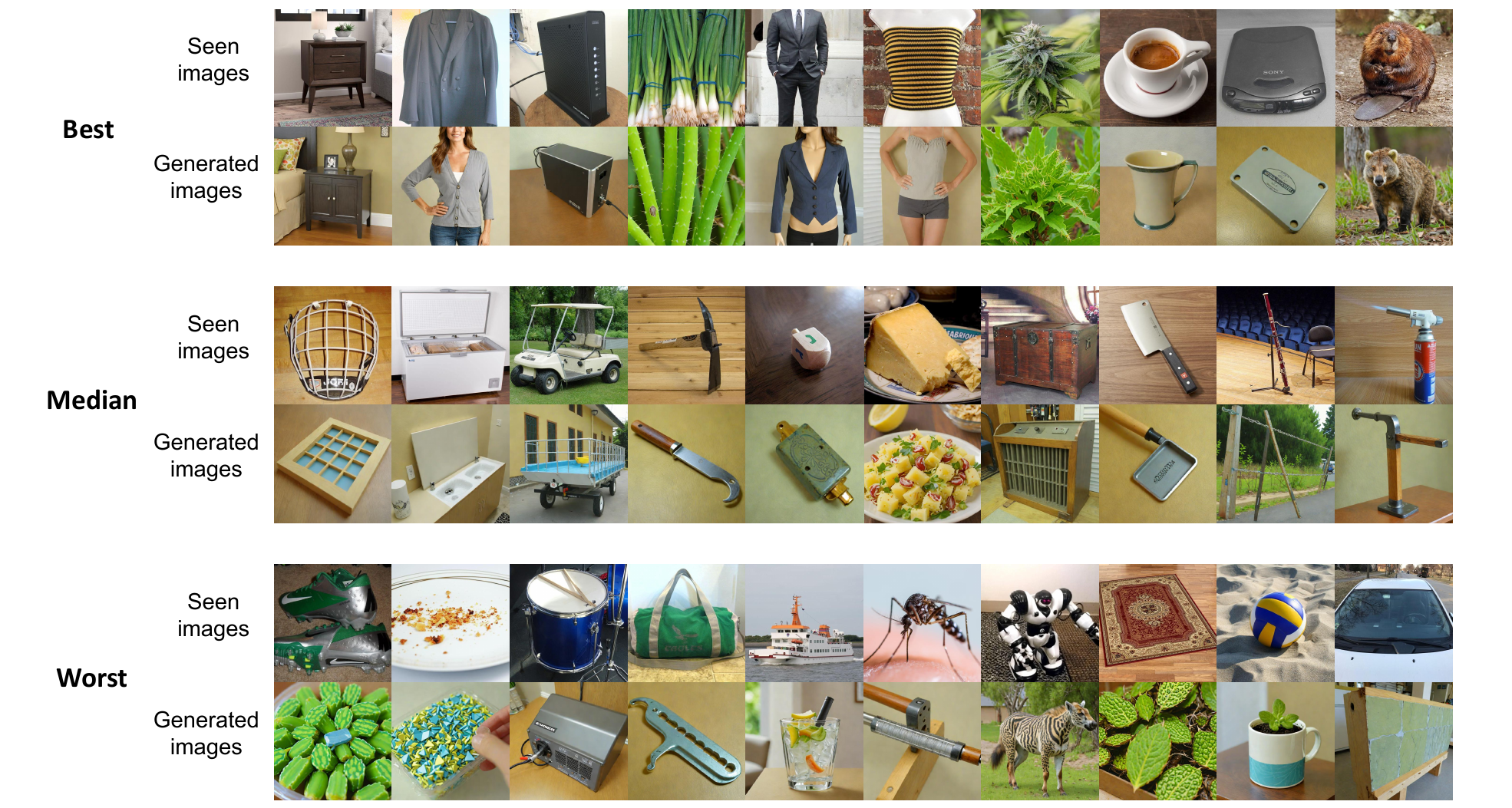}
    \caption{We compare the images reconstructed using UMind with the ground truth images, including those that show the best, median, and worst correspondence to the original stimulus images. }
\label{fig:best_median_worst}
\end{figure*}
%%%%%%%%%%%%%%%%%%%%%%%%%%%%%%%%%%%%%%%%%%%%%%%%
\begin{figure}
    \centering
    \includegraphics[width=1.0\linewidth]{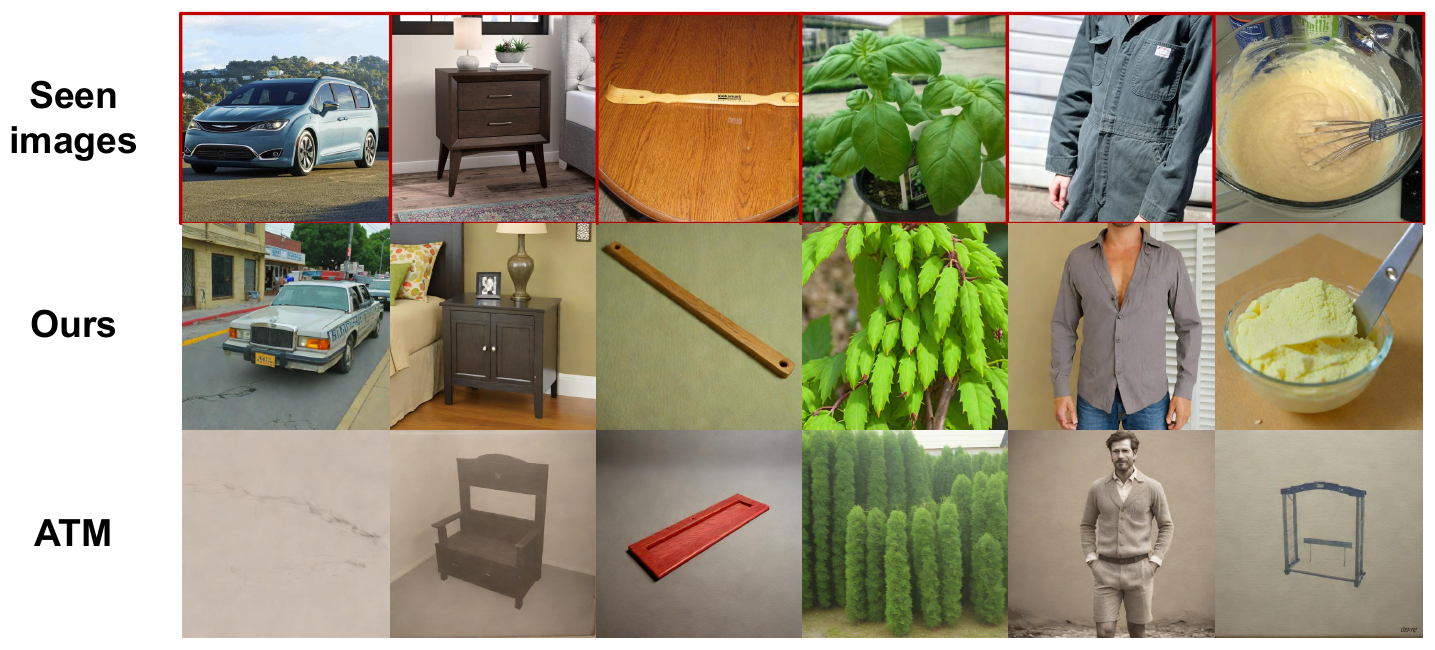}
    \caption{Comparison between the reconstructions using different methods.}
    \label{fig:images_comparison}
\end{figure}
%%%%%%%%%%%%%%%%%%%%%%%%%%%%%%%%%%%%%%%%%%%%%%%
\subsection{Reconstruction Performance}
\subsubsection{Quantitative Comparison}
% Although our model achieved excellent performance on retrieval and classification tasks, 
We further conducted image generation experiments on the THINGS-EEG and THINGS-MEG datasets to investigate its proficiency in learning neural-visual and semantic information.
We present the quantitative comparison between our model and recent state-of-the-art methods, including MB2C~\citep{weiMB2CMultimodalBidirectional2024}, B.D.~\citep{benchetritBrainDecodingRealtime2024}, and ATM~\citep{liVisualDecodingReconstruction2024}.
% We employ various high-level and low-level metrics to evaluate the similarity between the generated images and ground truth images.
% The metrics include PixCorr, which measures the pixel-wise correlation between the ground truth and generated images, and SSIM~\citep{wangImageQualityAssessment2004}, which evaluates structural similarity.
% EfficientNet-B1 (“EfficientNet”)~\citep{tanEfficientnetRethinkingModel2019} and SwAV-ResNet50 (“SwAV”)~\citep{caronUnsupervisedLearningVisual2020} are used to calculate the average correlation distance.
% All remaining metrics can be seen as a two-way retrieval task where the chance level is 50\%.
We employ various high-level and low-level metrics to evaluate the similarity between the generated images and ground truth images. Low-level metrics include pixel-level correlation (PixCorr), structural similarity index (SSIM), and the second and fifth layers of the AlexNet model (AlexNet(2) and AlexNet(5)), while high-level metrics encompass Inception, CLIP, EfficientNet, and SwAV-ResNet50 (SwAV) to calculate the visual and semantic fidelity.
For each category, we generate 10 images and rank them in descending order based on their similarity to the ground truth.
Finally, we compute the average metrics across these batches as our final results.

The experimental results are presented in Table~\ref{tab:generation_comparison}.
It can be observed that our UMind generates images that perform slightly lower on low-level metrics, such as PixCorr and SSIM, compared to other models on the THINGS-EEG dataset.
This is because methods like ATM leverage VAE latents to generate low-level images as the initial input for the diffusion model, thereby providing more low-level information.
% However, the images generated by our method outperform other state-of-the-art models significantly on high-level metrics.
However, our method significantly outperforms on high-level metrics. 
This improvement is attributed to the integration of dual-grained text, along with the guidance from aligned neural-visual representations, enabling the model to learn more detailed visual and semantic information.
On the THINGS-MEG dataset, our reconstruction performance surpasses that of ATM and achieves results comparable to B.D. framework.

\subsubsection{Qualitative Comparison}
We qualitatively compare the images reconstructed by our method with ground truth images.
As shown in Fig.~\ref{fig:best_median_worst}, the comparison includes the images that best, median, and worst correspond to the original stimulus images.
It can be observed that the images generated by our method are visually consistent with the ground truth images, including color, shape, and orientation.
Although some reconstructions do not effectively preserve the semantic information of the original images, they still capture detailed features such as similar colors and shapes.

As shown in Fig.~\ref{fig:images_comparison}, we also qualitatively compare the images reconstructed by our method with those by ATM. The images reconstructed using our method achieve greater visual and semantic consistency with the original images.
For example, in the second column of Fig.~\ref{fig:images_comparison}, our method accurately reconstructs the nightstand, along with details such as the bed beside it and the lamp on the nightstand.
In contrast, the image reconstructed by ATM only resembles a wooden box similar to a nightstand.

%%%%%%%%%%%%%%%%%%%%%%%%%%%%%%%%%%%%%%%%%%%%%%%%%%%%%
\begin{table*}[width=\textwidth, h]
    \centering
    % \captionsetup{font=small}
    \caption{Retrieval and classification accuracy (\%) with different EEG encoder on THINGS-EEG dataset}
    \setlength{\tabcolsep}{1.9pt}
    % \small
    % \footnotesize
    \scriptsize
    % \tiny
    \begin{tabular}{lcccccccccccccccccccccc}
        \toprule
        \multirow{2}*{Methods} & \multicolumn{2}{c}{Subject 1} & \multicolumn{2}{c}{Subject 2} & \multicolumn{2}{c}{Subject 3} & \multicolumn{2}{c}{Subject 4} & \multicolumn{2}{c}{Subject 5} & \multicolumn{2}{c}{Subject 6} & \multicolumn{2}{c}{Subject 7} & \multicolumn{2}{c}{Subject 8} & \multicolumn{2}{c}{Subject 9} & \multicolumn{2}{c}{Subject 10} & \multicolumn{2}{c}{Average} \\
        \cmidrule(l){2-3} \cmidrule(l){4-5} \cmidrule(l){6-7} \cmidrule(l){8-9} \cmidrule(l){10-11} \cmidrule(l){12-13} \cmidrule(l){14-15} \cmidrule(l){16-17} \cmidrule(l){18-19} \cmidrule(l){20-21} \cmidrule(l){22-23}
        & top-1 & top-5 & top-1 & top-5 & top-1 & top-5 & top-1 & top-5 & top-1 & top-5 & top-1 & top-5 & top-1 & top-5 & top-1 & top-5 & top-1 & top-5 & top-1 & top-5 & top-1 & top-5  \\
        \midrule
        \multicolumn{23}{c}{Retrieval accuracy} \\
        \midrule
        ShallowNet & 14.50 & 38.50 & 15.00 & 35.00 & 4.00 & 19.50 & 14.00 & 34.50 & 15.00 & 42.50 & 19.50 & 41.00 & 9.00 & 28.50 & 19.00 & 52.00 & 14.50 & 39.00 & 24.00 & 54.50 & 14.85 & 38.50 \\
        
        DeepNet & 12.00 & 34.00 & 16.50 & 40.00 & 17.50 & 45.00 & 21.00 & 51.00 & 11.50 & 38.00 & 18.00 & 46.50 & 16.00 & 41.00 & 24.50 & 56.50 & 18.00 & 44.00 & 25.00 & 52.00 & 18.00 & 44.80  \\
        
        EEGNet & 18.50 & 46.00 & 29.50 & 60.50 & 27.50 & 55.00 & 28.50 & 63.00 & 13.00 & 41.00 & 32.00 & 68.50 & 25.00 & 59.00 & 35.00 & 70.50 & 26.00 & 60.00 & 34.50 & 69.50 & 26.95 & 59.30 \\
        
        Conformer & 9.50 & 30.00 & 20.50 & 45.50 & 20.00 & 49.50 & 25.00 & 55.00 & 9.50 & 27.50 & 29.00 & 56.00 & 16.50 & 43.00 & 23.50 & 59.00 & 15.50 & 40.00 & 29.00 & 57.50 & 19.80 & 46.30 \\
        
        TSConv & 25.00 & 54.50 & 19.00 & 53.50 & 28.50 & 65.50 & 25.50 & 64.00 & 23.50 & 55.50 & 22.50 & 55.50 & 23.00 & 59.50 & 47.50 & 78.50 & 25.00 & 64.50 & 36.00 & 71.00 & 27.55 & 63.20 \\
        \rowcolor{blue!20} % Adjust the shade as needed
        
        ATM & 27.00 & 56.00 & 32.00 & 70.00 & 34.00 & 70.00 & 36.00 & 70.50 & 23.00 & 50.50 & 32.50 & 68.50 & 28.00 & 59.00 & 46.50 & 80.00 & 37.50 & 66.50 & 42.00 & 76.00 & \textbf{33.85} & \textbf{66.70} \\
        \midrule
        \multicolumn{23}{c}{Classification accuracy} \\
        \midrule
        ShallowNet & 4.00 & 14.00 & 1.00 & 8.00 & 1.00 & 4.50 & 4.00 & 13.00 & 5.50 & 23.00 & 1.00 & 9.00 & 2.50 & 10.50 & 4.50 & 18.50 & 5.00 & 16.00 & 6.00 & 19.00 & 3.45 & 13.55 \\
        
        DeepNet & 6.50 & 22.50 & 6.00 & 23.00 & 8.50 & 24.00 & 9.00 & 22.50 & 7.00 & 18.00 & 6.50 & 22.50 & 6.00 & 25.00 & 12.00 & 27.50 & 7.00 & 24.00 & 7.50 & 26.00 & 7.60 & 23.50  \\
        
        EEGNet & 5.00 & 27.50 & 8.50 & 27.00 & 9.50 & 30.00 & 12.50 & 33.50 & 8.00 & 21.50 & 10.50 & 31.00 & 9.50 & 30.50 & 16.50 & 35.50 & 8.50 & 25.00 & 12.00 & 37.00 & 10.05 & 29.85 \\
        
        Conformer & 4.50 & 17.50 & 8.00 & 21.50 & 6.00 & 28.50 & 10.50 & 30.00 & 5.00 & 15.00 & 8.50 & 25.00 & 5.50 & 24.50 & 9.50 & 23.50 & 7.00 & 22.00 & 11.00 & 30.50 & 7.55 & 23.80 \\
        
        TSConv & 9.50 & 30.00 & 7.00 & 23.00 & 10.00 & 31.50 & 13.50 & 29.50 & 6.50 & 22.50 & 6.00 & 23.50 & 9.50 & 35.50 & 15.50 & 36.50 & 10.00 & 35.00 & 14.00 & 36.50 & 10.15 & 30.35 \\
        \rowcolor{blue!20} % Adjust the shade as needed
        
        ATM & 7.50 & 31.00 & 7.50 & 32.00 & 15.50 & 37.50 & 17.00 & 38.50 & 8.00 & 22.50 & 10.50 & 37.50 & 11.50 & 30.00 & 16.00 & 46.00 & 10.00 & 32.50 & 16.50 & 36.50 & \textbf{12.00} & \textbf{34.40} \\
        \bottomrule
    \end{tabular}
    \label{tab:EEG_encoder_comparison}
\end{table*}
%%%%%%%%%%%%%%%%%%%%%%%%%%%%%%%%%%%%%%%%%%%%%%%%%%%%
\begin{table*}[width=\textwidth, htbp]
    \centering
    % \captionsetup{font=small}
    \caption{Retrieval and classification accuracy (\%) with different pre-trained CLIP on THINGS-EEG dataset}
    \setlength{\tabcolsep}{2pt}
    % \small
    % \footnotesize
    \scriptsize
    % \tiny
    \begin{tabular}{lcccccccccccccccccccccc}
        \toprule
        \multirow{2}*{Methods} & \multicolumn{2}{c}{Subject 1} & \multicolumn{2}{c}{Subject 2} & \multicolumn{2}{c}{Subject 3} & \multicolumn{2}{c}{Subject 4} & \multicolumn{2}{c}{Subject 5} & \multicolumn{2}{c}{Subject 6} & \multicolumn{2}{c}{Subject 7} & \multicolumn{2}{c}{Subject 8} & \multicolumn{2}{c}{Subject 9} & \multicolumn{2}{c}{Subject 10} & \multicolumn{2}{c}{Average} \\
        \cmidrule(l){2-3} \cmidrule(l){4-5} \cmidrule(l){6-7} \cmidrule(l){8-9} \cmidrule(l){10-11} \cmidrule(l){12-13} \cmidrule(l){14-15} \cmidrule(l){16-17} \cmidrule(l){18-19} \cmidrule(l){20-21} \cmidrule(l){22-23}
        & top-1 & top-5 & top-1 & top-5 & top-1 & top-5 & top-1 & top-5 & top-1 & top-5 & top-1 & top-5 & top-1 & top-5 & top-1 & top-5 & top-1 & top-5 & top-1 & top-5 & top-1 & top-5  \\
        \midrule
        \multicolumn{23}{c}{Retrieval accuracy} \\
        \midrule
        ViT-B/16 & 29.50 & 61.50 & 27.00 & 55.00 & 30.50 & 66.00 & 39.50 & 68.50 & 21.00 & 50.00 & 35.50 & 67.00 & 30.00 & 64.00 & 49.00 & 82.00 & 30.00 & 67.00 & 33.50 & 69.50 & 32.55 & 65.05 \\
        
        ViT-L/14 & 28.50 & 56.00 & 26.00 & 58.50 & 37.00 & 74.50 & 42.00 & 74.00 & 23.50 & 54.50 & 33.00 & 69.50 & 27.00 & 71.50 & 43.00 & 74.50 & 38.00 & 67.00 & 40.00 & 73.00 & 33.80 & 67.30 \\
        \rowcolor{blue!20} % Adjust the shade as needed
        ViT-H/14 & 27.00 & 56.00 & 32.00 & 70.00 & 34.00 & 70.00 & 36.00 & 70.50 & 23.00 & 50.50 & 32.50 & 68.50 & 28.00 & 59.00 & 46.50 & 80.00 & 37.50 & 66.50 & 42.00 & 76.00 & \textbf{33.85} & \textbf{66.70} \\
        
         ViT-G/14 & 25.50 & 51.50 & 33.00 & 65.50 & 23.00 & 52.50 & 31.50 & 68.50 & 20.00 & 48.50 & 25.50 & 58.00 & 28.00 & 60.00 & 48.00 & 77.50 & 18.00 & 46.00 & 39.50 & 68.50 & 29.20 & 59.65 \\
        \midrule
        \multicolumn{23}{c}{Classification accuracy} \\
        \midrule
        ViT-B/16 & 11.00 & 34.00 & 8.50 & 34.50 & 13.50 & 40.00 & 14.00 & 40.50 & 10.50 & 25.00 & 10.50 & 34.00 & 11.00 & 35.00 & 16.50 & 47.50 & 15.00 & 35.50 & 13.00 & 37.00 & 12.35 & 36.30 \\
        
        ViT-L/14 & 7.50 & 24.00 & 6.50 & 26.00 & 10.00 & 38.50 & 12.50 & 40.50 & 8.00 & 25.00 & 8.50 & 34.50 & 9.00 & 35.00 & 16.00 & 39.50 & 11.00 & 30.50 & 11.50 & 40.00 & 10.05 & 33.35  \\
        \rowcolor{blue!20} % Adjust the shade as needed
        ViT-H/14 & 7.50 & 31.00 & 7.50 & 32.00 & 15.50 & 37.50 & 17.00 & 38.50 & 8.00 & 22.50 & 10.50 & 37.50 & 11.50 & 30.00 & 16.00 & 46.00 & 10.00 & 32.50 & 16.50 & 36.50 & \textbf{12.00} & \textbf{34.40} \\

        ViT-G/14 & 8.00 & 23.50 & 7.50 & 29.00 & 9.50 & 23.50 & 13.50 & 42.00 & 8.00 & 28.00 & 11.50 & 31.50 & 11.00 & 32.00 & 15.00 & 37.50 & 9.00 & 22.50 & 11.00 & 36.50 & 10.40 & 30.60 \\
        \bottomrule
    \end{tabular}
    \label{tab:CLIP_comparison}
\end{table*}
%%%%%%%%%%%%%%%%%%%%%%%%%%%%%%%%%%%%%%%%%%%%%%%%%%%%%
\begin{table}[htbp]
    \centering
    % \captionsetup{font=small}
    \caption{Effects of dual grained text on retrieval \& classification (\%)}
    \setlength{\tabcolsep}{4pt}
    % \small
    % \footnotesize
    % \scriptsize
    % \tiny
    \begin{tabular}{cccccc}
        \toprule
        \multirow{2}*{coarse-grained} & \multirow{2}*{fine-grained} & \multicolumn{2}{c}{Retrieval} & \multicolumn{2}{c}{Classification} \\
        \cmidrule(l){3-4} \cmidrule(l){5-6} 
        & & top-1 & top-5 & top-1 & top-5 \\
        \midrule
        % \ding{55} & \ding{55} & \ding{55} & 18.65 & 40.50 & 0.40 & 2.50 \\

        \ding{55} & \ding{55} & 20.30 & 44.25 & 0.55 & 2.40 \\

        % \ding{55} & \checkmark & \ding{55} & 28.25 & 61.45 & 8.65 & 29.05 \\
        
        % \ding{55} & \ding{55} & \checkmark & 31.40 & 64.15 & 6.45 & 21.70 \\

        \checkmark & \ding{55} & 30.90 & 62.45 & 10.15 & 31.25 \\

        \ding{55} & \checkmark & 32.35 & 63.90 & 7.60 & 24.85 \\

        % \ding{55} & \checkmark & \checkmark & 33.15 & 64.50 & 10.75 & 31.10 \\

        % \rowcolor{blue!20} % Adjust the shade as needed
        \checkmark & \checkmark & \textbf{33.85} & \textbf{66.70} & \textbf{12.00} & \textbf{34.40} \\
        \bottomrule
    \end{tabular}
    \label{tab:ablation_decoding}
\end{table}

%%%%%%%%%%%%%%%%%%%%%%%%%%%%%%%%%%%%%%%%%%%%%%%
\begin{table*}[htbp]
    \centering
    % \captionsetup{font=small}
    \caption{Effects of coarse-grained text and fine-grained text on reconstruction performance.}
    \setlength{\tabcolsep}{3pt}
    % \small
    % \footnotesize
    % \scriptsize
    % \tiny
    \begin{tabular}{cccccccccc}
        \toprule
        \multirow{2}*{coarse-grained} & \multirow{2}*{fine-grained} & \multicolumn{4}{c}{Low-Level} & \multicolumn{4}{c}{High-Level} \\
        \cmidrule(lr){3-6} \cmidrule(l){7-10} 
         & & PixCorr $\uparrow$ & SSIM $\uparrow$ & AlexNet(2) $\uparrow$ & AlexNet(5) $\uparrow$ & Inception $\uparrow$ & CLIP $\uparrow$  & EfficientNet $\downarrow$ & SwAV $\downarrow$ \\
        \midrule
        \ding{55} & \checkmark & 0.143 & 0.398 & 0.669 & 0.779 & 0.686 & 0.756 & 0.901 & 0.583 \\
        
        \checkmark & \ding{55} & 0.148 & \textbf{0.399} & 0.689 & 0.805 & 0.709 & 0.779 & 0.899 & 0.584 \\
        
        \checkmark & \checkmark & \textbf{0.156} & 0.390 & \textbf{0.725} & \textbf{0.839} & \textbf{0.744} & \textbf{0.798} & \textbf{0.879} & \textbf{0.560} \\
        \bottomrule
    \end{tabular}
    \label{tab:ablation_generation}
\end{table*}
%%%%%%%%%%%%%%%%%%%%%%%%%%%%%%%%%%%%%%%%%%%%%%%%
\begin{table}[t]
    \centering
    % \captionsetup{font=small}
    \caption{Effect of loss function for retrieval and classification (\%)}
    \setlength{\tabcolsep}{8pt}
    % \small
    % \footnotesize
    % \scriptsize
    % \tiny
    \begin{tabular}{ccccc}
        \toprule
        \multirow{2}*{Loss} & \multicolumn{2}{c}{Retrieval} & \multicolumn{2}{c}{Classification} \\
        \cmidrule(l){2-3} \cmidrule(l){4-5} 
        & top-1 & top-5 & top-1 & top-5 \\
        \midrule
        w/o $\mathcal{L}_{CLIP}$ & 0.5 & 2.5 & 0.5 & 2.5 \\

        w/o $\mathcal{L}_{MSE}$ & 33.15 & 64.50 & 10.75 & 31.10 \\

        overall & \textbf{33.85} & \textbf{66.70} & \textbf{12.00} & \textbf{34.40} \\
        \bottomrule
    \end{tabular}
    \label{tab:ablation_decoding2}
\end{table}

%%%%%%%%%%%%%%%%%%%%%%%%%%%%%%%%%%%%%%%%%%%%%%%%%%%
\begin{table}[htbp]
    \centering
    % \captionsetup{font=small}
    \caption{Single task vs Multitask framework for retrieval \& classification (\%)}
    \setlength{\tabcolsep}{5pt}
    % \small
    % \footnotesize
    % \scriptsize
    % \tiny
    \begin{tabular}{cccccc}
        \toprule
        \multirow{2}*{Retrieval} & \multirow{2}*{Classification} & \multicolumn{2}{c}{Retrieval} & \multicolumn{2}{c}{Classification} \\
        \cmidrule(l){3-4} \cmidrule(l){5-6} 
        & & top-1 & top-5 & top-1 & top-5 \\
        \midrule
        \checkmark & \ding{55} & 31.85 & 64.15 & $-$ & $-$ \\

        \ding{55} & \checkmark & $-$ & $-$ & 10.50 & 30.75 \\

        \checkmark & \checkmark & \textbf{33.85} & \textbf{66.70} & \textbf{12.00} & \textbf{34.40} \\
        \bottomrule
    \end{tabular}
    \label{tab:ablation_multitask}
\end{table}
%%%%%%%%%%%%%%%%%%%%%%%%%%%%%%%%%%%%%%%%%%%%%%%

\subsection{Ablation Study}
\subsubsection{Effect of different M/EEG encoders}
% In recent years, researchers have proposed numerous powerful M/EEG encoders, we use the ATM~\citep{liVisualDecodingReconstruction2024} as our encoder to extract M/EEG features.
% It first employs a spatial self-attention mechanism to extract global spatial dependency features and then utilizes spatial-temporal convolution to further extract M/EEG spatial-temporal features for visual stimulus decoding and reconstruction.
% The image projector and text projector are both two-layer MLP used to map the EEG features into the CLIP image and CLIP text spaces respectively.
To investigate the impact of different encoders on the experimental results, we compare several representative M/EEG encoders, including EEGNet~\citep{lawhernEEGNetCompactConvolutional2018}, DeepConvNet, ShallowNet~\citep{schirrmeisterDeepLearningConvolutional2017}, Conformer~\citep{songEEGConformerConvolutional2022a}, TSConv in NICE~\citep{songDecodingNaturalImages2024}, and ATM~\citep{liVisualDecodingReconstruction2024}.
The comparison results of different encoders are shown in Table~\ref{tab:EEG_encoder_comparison}.
On the THINGS-EEG dataset, the ATM encoder outperforms other methods in both retrieval and classification tasks.
ATM achieves a top-1 accuracy of 33.85\% in the retrieval task and 12.00\% in the classification task, surpassing the best-performing TSConv by 6.30\% (\textit{p}\ \textless\ 0.01) and 1.85\% (\textit{p}\ \textless\ 0.05), respectively.
Therefore, we selected ATM as our M/EEG encoder.

\subsubsection{Effect of different image and text encoders}
% To facilitate the alignment of M/EEG features with both image features and text features, we perform multimodal alignment in a pre-aligned image-text space.
In the multimodal alignment module, we use the pre-trained image encoder and text encoder from Contrastive Language-Image Pre-training (CLIP)~\citep{radfordLearningTransferableVisual2021} to extract image and text features.
% To investigate the impact of different CLIP models on the decoding results, we employ various scales,
We systematically evaluate CLIP variants in different scales,
including OpenCLIP ViT-B/16, ViT-L/14, ViT-H/14, and ViT-G/14~\citep{ilharco_gabriel_2021,schuhmannLaion5bOpenLargescale2022}.
% All images and corresponding texts underwent standard preprocessing before training for direct feature extraction using CLIP.
The performance comparison of different pre-trained CLIP models is presented in Table~\ref{tab:CLIP_comparison}.
It can be observed that all pre-trained CLIP models demonstrate comparable performance in both retrieval and classification tasks.
Among them, OpenCLIP ViT-H/14 was chosen as our frozen image and text encoder because it consistently delivers strong results across both tasks.

\subsubsection{Effect of Coarse-grained and Fine-grained Text}
% In this study, we propose incorporating coarse-grained and fine-grained text for multimodal alignment, enabling the simultaneous learning of neural visual representations and neural semantic representations.
% To investigate the effects of coarse-grained text, fine-grained text, and MSE loss on visual stimulus retrieval and classification performance, we conducted ablation experiments on each component.
We conducted ablation experiments on the THINGS-EEG dataset to investigate the effects of coarse-grained and fine-grained text.
As presented in Table~\ref{tab:ablation_decoding},
% As observed, removing the MSE loss resulted in a decrease of 0.7\% (\textit{p}\ \textgreater\ 0.05) and 1.25\% (\textit{p}\ \textless\ 0.05) in top-1 accuracy for retrieval and classification, respectively.
% This indicates that incorporating an appropriately weighted MSE loss for the subsequent reconstruction task can also enhance the model's performance in retrieval and classification.
without text modality data, the framework achieves top-1 classification accuracies of only 0.55\%, indicating that images provide very limited semantic information for M/EEG visual stimulus classification.
After incorporating coarse-grained and fine-grained text, the model's classification performance improves significantly, and retrieval accuracy is also notably enhanced.
This demonstrates that introducing the text modality for multimodal alignment not only enables the extraction of detailed semantic information but also enhances the extraction of neural visual representations.

Table~\ref{tab:ablation_generation} also presented the effects of dual-grained text on visual stimulus reconstruction.
% We extracted ground truth prompt embeddings using either coarse-grained or fine-grained text exclusively.
% The results show that when using only coarse-grained or only fine-grained text, the quality of the reconstructed images declines.
The results indicate that relying on only one type of text granularity significantly reduces the quality of reconstructed images compared to using both granularity levels together. 
% The categorical information in coarse-grained text and the detailed descriptions in fine-grained text complement each other.
Specifically, the model achieves better reconstruction performance when both coarse and fine-grained text are employed, as their complementary information allows the model to capture more semantic details.
Therefore, our model can learn semantically rich neural representations and reconstruct images that are semantically similar to the ground truth visual stimuli.

\subsubsection{Effect of Different Loss Functions}
Table~\ref{tab:ablation_decoding2} shows the effect of different loss functions.
As observed, removing the MSE loss resulted in a decrease of 0.7\% (\textit{p}\ \textgreater\ 0.05) and 1.25\% (\textit{p}\ \textless\ 0.05) in top-1 accuracy for retrieval and classification, respectively.
This indicates that incorporating an appropriately weighted MSE loss for the subsequent reconstruction task can also enhance the model's performance in retrieval and classification.
Using only MSE loss is ineffective for multimodal alignment, whereas combining contrastive learning with MSE loss enhances retrieval and classification performance.

\subsubsection{Effect of Multitask Framework}
To investigate whether our multitask framework can facilitate mutual enhancement among different tasks, we trained separate single-task models using only image and text data for M/EEG visual retrieval and classification, respectively.
As shown in Table~\ref{tab:ablation_multitask}, the results demonstrate that our unified multitask framework outperforms the single-task retrieval and classification frameworks by 2.00\% (\textit{p}\ \textgreater\ 0.05) and 1.50\% (\textit{p}\ \textless\ 0.05) in top-1 accuracy, respectively.
This suggests that incorporating text for multimodal alignment not only enables the model to handle different tasks but also facilitates mutual enhancement between tasks, leading to more effective learning of neural visual and semantic representations.

%%%%%%%%%%%%%%%%%%%%%%%%%%%%%%%%%%%%%%%%%%%%%%%
\begin{figure}
    \centering
    \includegraphics[width=1.0\linewidth]{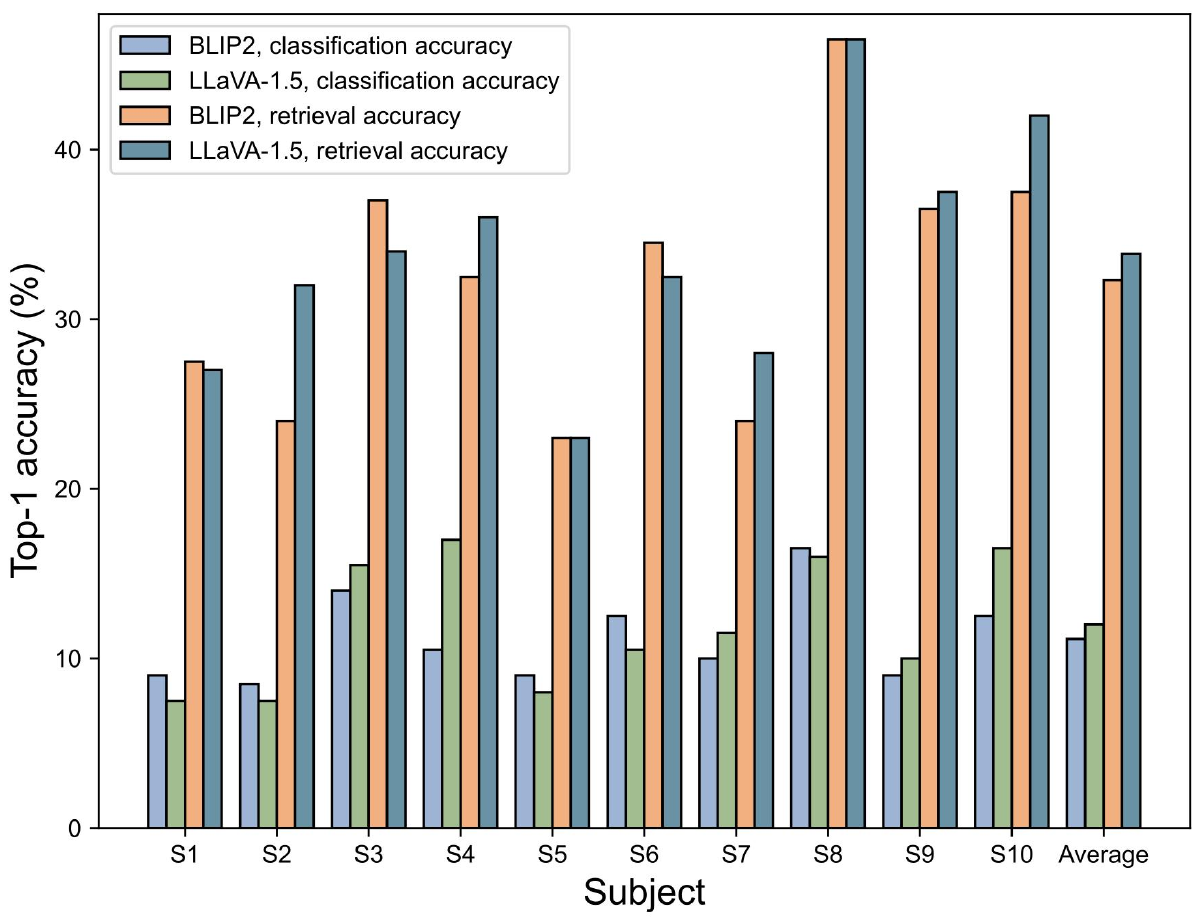}
    \caption{Comparison of results for visual stimulus retrieval and classification using fine-grained text generated by LLaVA-1.5 and BLIP2.}
    \label{fig:LLaVA_BLIP2_comparison}
\end{figure}
%%%%%%%%%%%%%%%%%%%%%%%%%%%%%%%%%%%%%%%%%%%%%%%
\subsubsection{Effect of Different Image-to-Text Generation Models}
To assess the impact of fine-grained text generated by different pre-trained image-to-text models on experimental results, we conducted a comparative analysis using two models: LLaVA-1.5~\citep{liuImprovedBaselinesVisual2024} and BLIP2~\citep{liBlip2BootstrappingLanguageimage2023}.
The results are shown in Fig.~\ref{fig:LLaVA_BLIP2_comparison}.
It can be observed that for both retrieval and classification tasks, the average top-1 accuracy of using fine-grained text generated by LLaVA-1.5 surpasses that of BLIP2.
This is because the image captions generated by BLIP2 are often short and simplistic, such as ``the antelope is brown."
In contrast, the text generated by LLaVA-1.5 contains richer details about the image, including information on location, color, and direction, as illustrated in Fig.~\ref{fig:coarse_fine_comparison}.
Consequently, the fine-grained text generated by LLaVA-1.5 better facilitates the model in learning neural semantic representations with more detailed information.

        %% 6.Discussion
	\section{Discussion}
        \label{sec:Discussion}
Current approaches for decoding visual stimuli primarily rely on contrastive learning to align brain signals with images, overlooking the critical role of the textual modality.
Additionally, these models are often limited to addressing a single task or require training a separate encoder for each task.
To address these limitations, we propose a unified multitask framework UMind capable of simultaneously performing zero-shot visual stimulus retrieval, classification, and reconstruction tasks.
% The coarse-grained text provides only categorical information while lacking detailed descriptions of visual stimuli, such as color, shape, and background.
% In contrast, fine-grained text generated by pre-trained image-to-text models includes more detailed information, but the categorical descriptions of visual stimuli may sometimes be unclear or even inaccurate.
By leveraging both coarse-grained and fine-grained text for multimodal alignment, our framework facilitates the learning of neural-visual and neural-semantic representations enriched with finer details.
This enhances the extraction of neural representations that are beneficial for visual stimulus retrieval and classification tasks.
% Consequently, coarse-grained and fine-grained text can complement each other, enabling the model to learn more effective neural semantic representations.
% By simultaneously incorporating both coarse-grained and fine-grained text for multimodal alignment, our approach not only facilitates the learning of neural semantic representations enriched with finer details but also enhances the extraction of neural visual representations.
% The extracted neural visual and semantic representations can be utilized for visual stimulus retrieval and classification tasks, respectively.
Furthermore, these extracted neural representations serve as dual conditioning inputs for the pre-trained diffusion model, guiding the generation of realistic and semantically meaningful images from both visual and semantic perspectives.

Nevertheless, there are some limitations in our work.
By incorporating coarse-grained and fine-grained text, we have successfully learned more detailed neural semantic representations associated with visual stimuli, which are effective for classification and reconstruction tasks.
However, decoding text corresponding to visual stimuli from these semantic representations remains a significant challenge.
Additionally, although we have reconstructed images from M/EEG signals, the quality of the reconstructed images still falls short compared to fMRI-based works.
Further research is needed to narrow this gap.

        %% 7.Conclusion
	\section{Conclusion}
        \label{sec:Conclusion}
In this work, we propose \textbf{UMind}, a novel multitask multimodal alignment framework that simultaneously addresses zero-shot M/EEG-based visual stimulus retrieval, classification, and reconstruction tasks.
By incorporating both coarse-grained and fine-grained text, we can not only extract more detailed neural semantic representations but also enhance the learning of neural visual representations.
The extracted neural visual and semantic representations can be used for M/EEG visual stimulus retrieval and classification, respectively.
Additionally, they serve as dual conditional inputs to a frozen diffusion model to guide image generation.
Extensive quantitative and qualitative experiments demonstrate that our framework achieves state-of-the-art performance across two datasets, highlighting its effectiveness in learning robust visual and semantic representations for visual stimulus decoding.
% This indicates that our framework effectively learns visual and semantic representations for visual stimulus decoding.

	\section*{Declaration of generative AI and AI-assisted technologies in the writing process}
	\par{During the preparation of this work, the authors used ChatGPT to improve language and readability.   After using this tool, the authors reviewed and edited the content as needed and take full responsibility for the content of the publication.}

	\bibliographystyle{model5-names}
	% \bibliographystyle{cas-model2-names}

	\bibliography{ref}

@article{allenMassive7TFMRI2022,
  title = {A Massive {{7T fMRI}} Dataset to Bridge Cognitive Neuroscience and Artificial Intelligence},
  author = {Allen, Emily J. and {St-Yves}, Ghislain and Wu, Yihan and Breedlove, Jesse L. and Prince, Jacob S. and Dowdle, Logan T. and Nau, Matthias and Caron, Brad and Pestilli, Franco and Charest, Ian},
  year = {2022},
  journal = {Nature neuroscience},
  volume = {25},
  number = {1},
  pages = {116--126},
  publisher = {Nature Publishing Group US New York}
}

@misc{benchetritBrainDecodingRealtime2024,
  title = {Brain Decoding: Toward Real-Time Reconstruction of Visual Perception},
  author={Yohann Benchetrit and Hubert Banville and Jean-Rémi King},
  year={2024},
  eprint={2310.19812},
  archivePrefix={arXiv},
  primaryClass={eess.IV},
  url={https://arxiv.org/abs/2310.19812}, 
}

@misc{fuBrainVisExploringBridge2024,
  title = {{{BrainVis}}: {{Exploring}} the {{Bridge}} between {{Brain}} and {{Visual Signals}} via {{Image Reconstruction}}},
  shorttitle = {{{BrainVis}}},
  author={Honghao Fu and Zhiqi Shen and Jing Jih Chin and Hao Wang},
  year={2024},
  eprint={2312.14871},
  archivePrefix={arXiv},
  primaryClass={cs.CV},
  url={https://arxiv.org/abs/2312.14871},
}

@inproceedings{jiaoDecodingEEGVisualguided2019,
  title = {Decoding {{EEG}} by {{Visual-guided Deep Neural Networks}}.},
  booktitle = {{{IJCAI}}},
  author = {Jiao, Zhicheng and You, Haoxuan and Yang, Fan and Li, Xin and Zhang, Han and Shen, Dinggang},
  year = {2019},
  volume = {28},
  pages = {1387--1393},
  publisher = {Macao}
}

@article{kayIdentifyingNaturalImages2008,
  title = {Identifying Natural Images from Human Brain Activity},
  author = {Kay, Kendrick N. and Naselaris, Thomas and Prenger, Ryan J. and Gallant, Jack L.},
  year = {2008},
  journal = {Nature},
  volume = {452},
  number = {7185},
  pages = {352--355},
  publisher = {Nature Publishing Group UK London}
}

@inproceedings{liBlip2BootstrappingLanguageimage2023,
  title = {Blip-2: {{Bootstrapping}} Language-Image Pre-Training with Frozen Image Encoders and Large Language Models},
  shorttitle = {Blip-2},
  booktitle = {International Conference on Machine Learning},
  author = {Li, Junnan and Li, Dongxu and Savarese, Silvio and Hoi, Steven},
  year = {2023},
  pages = {19730--19742},
  publisher = {PMLR}
}

@inproceedings{linMindReaderReconstructing2022,
  title = {Mind Reader: {{Reconstructing}} Complex Images from Brain Activities},
  shorttitle = {Mind Reader},
  author = {Lin, Sikun and Sprague, Thomas and Singh, Ambuj K.},
  year = {2022},
  booktitle = {Advances in Neural Information Processing Systems},
  volume = {35},
  pages = {29624--29636}
}

@inproceedings{liVisualDecodingReconstruction2024,
 title = {Visual Decoding and Reconstruction via {EEG} Embeddings with Guided Diffusion},
 author = {Li, Dongyang and Wei, Chen and Li, Shiying and Zou, Jiachen and Liu, Quanying},
 booktitle = {Advances in Neural Information Processing Systems},
 volume = {37},
 pages = {102822--102864},
 year = {2024}
}

@article{miyawakiVisualImageReconstruction2008,
  title = {Visual Image Reconstruction from Human Brain Activity Using a Combination of Multiscale Local Image Decoders},
  author = {Miyawaki, Yoichi and Uchida, Hajime and Yamashita, Okito and Sato, Masa-aki and Morito, Yusuke and Tanabe, Hiroki C. and Sadato, Norihiro and Kamitani, Yukiyasu},
  year = {2008},
  journal = {Neuron},
  volume = {60},
  number = {5},
  pages = {915--929},
  publisher = {Elsevier}
}

@article{nishimotoReconstructingVisualExperiences2011,
  title = {Reconstructing Visual Experiences from Brain Activity Evoked by Natural Movies},
  author = {Nishimoto, Shinji and Vu, An T. and Naselaris, Thomas and Benjamini, Yuval and Yu, Bin and Gallant, Jack L.},
  year = {2011},
  journal = {Current biology},
  volume = {21},
  number = {19},
  pages = {1641--1646},
  publisher = {Elsevier}
}

@article{rameshHierarchicalTextconditionalImage2022,
  title = {Hierarchical Text-Conditional Image Generation with Clip Latents},
  author = {Ramesh, Aditya and Dhariwal, Prafulla and Nichol, Alex and Chu, Casey and Chen, Mark},
  year = {2022},
  journal = {arXiv preprint arXiv:2204.06125},
  volume = {1},
  number = {2},
  eprint = {2204.06125},
  pages = {3},
  archiveprefix = {arXiv}
}

@article{songEEGConformerConvolutional2022a,
  title = {{{EEG}} Conformer: {{Convolutional}} Transformer for {{EEG}} Decoding and Visualization},
  shorttitle = {{{EEG}} Conformer},
  author = {Song, Yonghao and Zheng, Qingqing and Liu, Bingchuan and Gao, Xiaorong},
  year = {2022},
  journal = {IEEE Transactions on Neural Systems and Rehabilitation Engineering},
  volume = {31},
  pages = {710--719},
  publisher = {IEEE}
}

@inproceedings{spampinatoDeepLearningHuman2017,
  title = {Deep Learning Human Mind for Automated Visual Classification},
  booktitle = {Proceedings of the {{IEEE}} Conference on Computer Vision and Pattern Recognition},
  author = {Spampinato, Concetto and Palazzo, Simone and Kavasidis, Isaak and Giordano, Daniela and Souly, Nasim and Shah, Mubarak},
  year = {2017},
  pages = {6809--6817}
}

@inproceedings{takagiHighresolutionImageReconstruction2023,
  title = {High-Resolution Image Reconstruction with Latent Diffusion Models from Human Brain Activity},
  booktitle = {Proceedings of the {{IEEE}}/{{CVF Conference}} on {{Computer Vision}} and {{Pattern Recognition}}},
  author = {Takagi, Yu and Nishimoto, Shinji},
  year = {2023},
  pages = {14453--14463}
}

@inproceedings{weiMB2CMultimodalBidirectional2024,
  title = {{{MB2C}}: {{Multimodal Bidirectional Cycle Consistency}} for {{Learning Robust Visual Neural Representations}}},
  shorttitle = {{{MB2C}}},
  booktitle = {{{ACM Multimedia}} 2024},
  author = {Wei, Yayun and Cao, Lei and Li, Hao and Dong, Yilin},
  year = {2024}
}

@article{wenNeuralEncodingDecoding2018,
  title = {Neural Encoding and Decoding with Deep Learning for Dynamic Natural Vision},
  author = {Wen, Haiguang and Shi, Junxing and Zhang, Yizhen and Lu, Kun-Han and Cao, Jiayue and Liu, Zhongming},
  year = {2018},
  journal = {Cerebral cortex},
  volume = {28},
  number = {12},
  pages = {4136--4160},
  publisher = {Oxford University Press}
}

@article{renReconstructingSeenImage2021,
  title = {Reconstructing Seen Image from Brain Activity by Visually-Guided Cognitive Representation and Adversarial Learning},
  author = {Ren, Ziqi and Li, Jie and Xue, Xuetong and Li, Xin and Yang, Fan and Jiao, Zhicheng and Gao, Xinbo},
  year = {2021},
  journal = {NeuroImage},
  volume = {228},
  pages = {117602},
  publisher = {Elsevier}
}

@inproceedings{ozcelikReconstructionPerceivedImages2022,
  title = {Reconstruction of Perceived Images from Fmri Patterns and Semantic Brain Exploration Using Instance-Conditioned Gans},
  booktitle = {2022 {{International Joint Conference}} on {{Neural Networks}} ({{IJCNN}})},
  author = {Ozcelik, Furkan and Choksi, Bhavin and Mozafari, Milad and Reddy, Leila and VanRullen, Rufin},
  year = {2022},
  pages = {1--8},
  publisher = {IEEE}
}

@inproceedings{goodfellowGenerativeAdversarialNets2014,
  title = {Generative Adversarial Nets},
  author = {Goodfellow, Ian and {Pouget-Abadie}, Jean and Mirza, Mehdi and Xu, Bing and {Warde-Farley}, David and Ozair, Sherjil and Courville, Aaron and Bengio, Yoshua},
  year = {2014},
  booktitle = {Advances in neural information processing systems},
  volume = {27}
}

@inproceedings{casanovaInstanceconditionedGan2021,
  title = {Instance-Conditioned Gan},
  author = {Casanova, Arantxa and Careil, Marlene and Verbeek, Jakob and Drozdzal, Michal and Romero Soriano, Adriana},
  year = {2021},
  booktitle = {Advances in Neural Information Processing Systems},
  volume = {34},
  pages = {27517--27529}
}

@inproceedings{rombachHighresolutionImageSynthesis2022,
  title = {High-Resolution Image Synthesis with Latent Diffusion Models},
  booktitle = {Proceedings of the {{IEEE}}/{{CVF}} Conference on Computer Vision and Pattern Recognition},
  author = {Rombach, Robin and Blattmann, Andreas and Lorenz, Dominik and Esser, Patrick and Ommer, Bj{\"o}rn},
  year = {2022},
  pages = {10684--10695}
}

@inproceedings{chenSeeingBrainConditional2023,
  title = {Seeing beyond the Brain: {{Conditional}} Diffusion Model with Sparse Masked Modeling for Vision Decoding},
  shorttitle = {Seeing beyond the Brain},
  booktitle = {Proceedings of the {{IEEE}}/{{CVF Conference}} on {{Computer Vision}} and {{Pattern Recognition}}},
  author = {Chen, Zijiao and Qing, Jiaxin and Xiang, Tiange and Yue, Wan Lin and Zhou, Juan Helen},
  year = {2023},
  pages = {22710--22720}
}

@article{ozcelikNaturalSceneReconstruction2023,
  title = {Natural Scene Reconstruction from {{fMRI}} Signals Using Generative Latent Diffusion},
  author = {Ozcelik, Furkan and VanRullen, Rufin},
  year = {2023},
  journal = {Scientific Reports},
  volume = {13},
  number = {1},
  pages = {15666},
  publisher = {Nature Publishing Group UK London}
}

@inproceedings{xuVersatileDiffusionText2023,
  title = {Versatile Diffusion: {{Text}}, Images and Variations All in One Diffusion Model},
  shorttitle = {Versatile Diffusion},
  booktitle = {Proceedings of the {{IEEE}}/{{CVF International Conference}} on {{Computer Vision}}},
  author = {Xu, Xingqian and Wang, Zhangyang and Zhang, Gong and Wang, Kai and Shi, Humphrey},
  year = {2023},
  pages = {7754--7765}
}

@inproceedings{scottiReconstructingMindsEye2024,
  title = {Reconstructing the Mind's Eye: {{fMRI-to-image}} with Contrastive Learning and Diffusion Priors},
  shorttitle = {Reconstructing the Mind's Eye},
  author = {Scotti, Paul and Banerjee, Atmadeep and Goode, Jimmie and Shabalin, Stepan and Nguyen, Alex and Dempster, Aidan and Verlinde, Nathalie and Yundler, Elad and Weisberg, David and Norman, Kenneth},
  year = {2024},
  booktitle = {Advances in Neural Information Processing Systems},
  volume = {36}
}

@misc{scottiMindEye2SharedSubjectModels2024,
  title = {{{MindEye2}}: {{Shared-Subject Models Enable fMRI-To-Image With}} 1 {{Hour}} of {{Data}}},
  shorttitle = {{{MindEye2}}},
  author={Paul S. Scotti and Mihir Tripathy and Cesar Kadir Torrico Villanueva and Reese Kneeland and Tong Chen and Ashutosh Narang and Charan Santhirasegaran and Jonathan Xu and Thomas Naselaris and Kenneth A. Norman and Tanishq Mathew Abraham},
  year={2024},
  eprint={2403.11207},
  archivePrefix={arXiv},
  primaryClass={cs.CV},
  url={https://arxiv.org/abs/2403.11207}, 
}

@inproceedings{spampinatoDeepLearningHuman2017a,
  title = {Deep Learning Human Mind for Automated Visual Classification},
  booktitle = {Proceedings of the {{IEEE}} Conference on Computer Vision and Pattern Recognition},
  author = {Spampinato, Concetto and Palazzo, Simone and Kavasidis, Isaak and Giordano, Daniela and Souly, Nasim and Shah, Mubarak},
  year = {2017},
  pages = {6809--6817}
}

@inproceedings{palazzoGenerativeAdversarialNetworks2017,
  title = {Generative Adversarial Networks Conditioned by Brain Signals},
  booktitle = {Proceedings of the {{IEEE}} International Conference on Computer Vision},
  author = {Palazzo, Simone and Spampinato, Concetto and Kavasidis, Isaak and Giordano, Daniela and Shah, Mubarak},
  year = {2017},
  pages = {3410--3418}
}

@inproceedings{kavasidisBrain2ImageConvertingBrain2017,
  title = {{{{\emph{Brain2Image}}}}: {{Converting Brain Signals}} into {{Images}}},
  shorttitle = {{{{\emph{Brain2Image}}}}},
  booktitle = {Proceedings of the 25th {{ACM}} International Conference on {{Multimedia}}},
  author = {Kavasidis, Isaak and Palazzo, Simone and Spampinato, Concetto and Giordano, Daniela and Shah, Mubarak},
  year = {2017},
  pages = {1809--1817},
  publisher = {ACM},
  address = {Mountain View California USA},
  doi = {10.1145/3123266.3127907},
  isbn = {978-1-4503-4906-2}
}

@inproceedings{tirupatturThoughtVizVisualizingHuman2018,
  title = {{{ThoughtViz}}: {{Visualizing Human Thoughts Using Generative Adversarial Network}}},
  shorttitle = {{{ThoughtViz}}},
  booktitle = {Proceedings of the 26th {{ACM}} International Conference on {{Multimedia}}},
  author = {Tirupattur, Praveen and Rawat, Yogesh Singh and Spampinato, Concetto and Shah, Mubarak},
  year = {2018},
  pages = {950--958},
  publisher = {ACM},
  doi = {10.1145/3240508.3240641},
  isbn = {978-1-4503-5665-7}
}

@article{liPerilsPitfallsBlock2020,
  title = {The Perils and Pitfalls of Block Design for {{EEG}} Classification Experiments},
  author = {Li, Ren and Johansen, Jared S. and Ahmed, Hamad and Ilyevsky, Thomas V. and Wilbur, Ronnie B. and Bharadwaj, Hari M. and Siskind, Jeffrey Mark},
  year = {2020},
  journal = {IEEE Transactions on Pattern Analysis and Machine Intelligence},
  volume = {43},
  number = {1},
  pages = {316--333},
  publisher = {IEEE}
}

@article{giffordLargeRichEEG2022,
  title = {A Large and Rich {{EEG}} Dataset for Modeling Human Visual Object Recognition},
  author = {Gifford, Alessandro T. and Dwivedi, Kshitij and Roig, Gemma and Cichy, Radoslaw M.},
  year = {2022},
  journal = {NeuroImage},
  volume = {264},
  pages = {119754},
  publisher = {Elsevier}
}

@inproceedings{karrasAnalyzingImprovingImage2020,
  title = {Analyzing and Improving the Image Quality of Stylegan},
  booktitle = {Proceedings of the {{IEEE}}/{{CVF}} Conference on Computer Vision and Pattern Recognition},
  author = {Karras, Tero and Laine, Samuli and Aittala, Miika and Hellsten, Janne and Lehtinen, Jaakko and Aila, Timo},
  year = {2020},
  pages = {8110--8119}
}

@inproceedings{liuImprovedBaselinesVisual2024,
  title = {Improved Baselines with Visual Instruction Tuning},
  booktitle = {Proceedings of the {{IEEE}}/{{CVF Conference}} on {{Computer Vision}} and {{Pattern Recognition}}},
  author = {Liu, Haotian and Li, Chunyuan and Li, Yuheng and Lee, Yong Jae},
  year = {2024},
  pages = {26296--26306}
}

@inproceedings{radfordLearningTransferableVisual2021,
  title = {Learning Transferable Visual Models from Natural Language Supervision},
  booktitle = {International Conference on Machine Learning},
  author = {Radford, Alec and Kim, Jong Wook and Hallacy, Chris and Ramesh, Aditya and Goh, Gabriel and Agarwal, Sandhini and Sastry, Girish and Askell, Amanda and Mishkin, Pamela and Clark, Jack},
  year = {2021},
  pages = {8748--8763},
  publisher = {PMLR}
}

@inproceedings{schuhmannLaion5bOpenLargescale2022,
  title = {Laion-5b: {{An}} Open Large-Scale Dataset for Training next Generation Image-Text Models},
  shorttitle = {Laion-5b},
  author = {Schuhmann, Christoph and Beaumont, Romain and Vencu, Richard and Gordon, Cade and Wightman, Ross and Cherti, Mehdi and Coombes, Theo and Katta, Aarush and Mullis, Clayton and Wortsman, Mitchell},
  year = {2022},
  booktitle = {Advances in Neural Information Processing Systems},
  volume = {35},
  pages = {25278--25294}
}

@software{ilharco_gabriel_2021,
  author = {Ilharco, Gabriel and Wortsman, Mitchell and Wightman, Ross and Gordon, Cade and Carlini, Nicholas and Taori, Rohan and Dave, Achal and Shankar, Vaishaal and Namkoong, Hongseok and Miller, John and Hajishirzi, Hannaneh and Farhadi, Ali and Schmidt, Ludwig},
  title = {OpenCLIP},
  year = 2021,
  publisher = {Zenodo},
  version  = {0.1},
  doi = {10.5281/zenodo.5143773},
  url = {https://doi.org/10.5281/zenodo.5143773}
}

@article{schirrmeisterDeepLearningConvolutional2017,
  title = {Deep Learning with Convolutional Neural Networks for {{EEG}} Decoding and Visualization},
  author = {Schirrmeister, Robin Tibor and Springenberg, Jost Tobias and Fiederer, Lukas Dominique Josef and Glasstetter, Martin and Eggensperger, Katharina and Tangermann, Michael and Hutter, Frank and Burgard, Wolfram and Ball, Tonio},
  year = {2017},
  journal = {Human Brain Mapping},
  volume = {38},
  number = {11},
  pages = {5391--5420},
  doi = {10.1002/hbm.23730},
  copyright = {http://creativecommons.org/licenses/by/4.0/}
}

@article{lawhernEEGNetCompactConvolutional2018,
  title = {{{EEGNet}}: A Compact Convolutional Neural Network for {{EEG-based}} Brain--Computer Interfaces},
  shorttitle = {{{EEGNet}}},
  author = {Lawhern, Vernon J. and Solon, Amelia J. and Waytowich, Nicholas R. and Gordon, Stephen M. and Hung, Chou P. and Lance, Brent J.},
  year = {2018},
  journal = {Journal of neural engineering},
  volume = {15},
  number = {5},
  pages = {056013},
  publisher = {iOP Publishing}
}

@inproceedings{sauerAdversarialDiffusionDistillation2025,
  title = {Adversarial {{Diffusion Distillation}}},
  booktitle = {{European} {Conference} on {Computer} {Vision}},
  author = {Sauer, Axel and Lorenz, Dominik and Blattmann, Andreas and Rombach, Robin},
  year = {2025},
  volume = {15144},
  pages = {87--103},
  isbn = {978-3-031-73015-3 978-3-031-73016-0}
}

@misc{ye2023ipadaptertextcompatibleimage,
      title = {{{IP-Adapter}}: {{Text Compatible Image Prompt Adapter}} for {{Text-to-Image Diffusion Models}}}, 
      shorttitle = {{{IP-Adapter}}},
      author = {Ye, Hu and Zhang, Jun and Liu, Sibo and Han, Xiao and Yang, Wei},
      year = {2023},
      number = {arXiv:2308.06721},
      eprint = {2308.06721},
      publisher = {arXiv},
      doi = {10.48550/arXiv.2308.06721},
      archiveprefix = {arXiv} 
}

@inproceedings{vaswaniAttentionAllYou2017,
  author = {Vaswani, Ashish and Shazeer, Noam and Parmar, Niki and Uszkoreit, Jakob and Jones, Llion and Gomez, Aidan N and Kaiser, \L ukasz and Polosukhin, Illia},
  booktitle = {Advances in Neural Information Processing Systems},
  publisher = {Curran Associates, Inc.},
  title = {{Attention} is {All} you {Need}},
  volume = {30},
  year = {2017}
}

@article{guggenmosMultivariatePatternAnalysis2018,
  title = {Multivariate Pattern Analysis for {{MEG}}: {{A}} Comparison of Dissimilarity Measures},
  shorttitle = {Multivariate Pattern Analysis for {{MEG}}},
  author = {Guggenmos, Matthias and Sterzer, Philipp and Cichy, Radoslaw Martin},
  year = {2018},
  journal = {Neuroimage},
  volume = {173},
  pages = {434--447},
  publisher = {Elsevier}
}

@article{hebartTHINGSdataMultimodalCollection2023,
  title = {{{THINGS-data}}, a Multimodal Collection of Large-Scale Datasets for Investigating Object Representations in Human Brain and Behavior},
  author = {Hebart, Martin N. and Contier, Oliver and Teichmann, Lina and Rockter, Adam H. and Zheng, Charles Y. and Kidder, Alexis and Corriveau, Anna and {Vaziri-Pashkam}, Maryam and Baker, Chris I.},
  year = {2023},
  journal = {Elife},
  volume = {12},
  pages = {e82580},
  publisher = {eLife Sciences Publications Limited}
}

@article{duDecodingVisualNeural2023,
  title = {Decoding Visual Neural Representations by Multimodal Learning of Brain-Visual-Linguistic Features},
  author = {Du, Changde and Fu, Kaicheng and Li, Jinpeng and He, Huiguang},
  year = {2023},
  journal = {IEEE Transactions on Pattern Analysis and Machine Intelligence},
  volume = {45},
  number = {9},
  pages = {10760--10777},
  publisher = {IEEE}
}

@article{jiangBrainmediaDeepFramework2020,
  title = {A Brain-Media Deep Framework towards Seeing Imaginations inside Brains},
  author = {Jiang, Jianmin and Fares, Ahmed and Zhong, Sheng-Hua},
  year = {2020},
  journal = {IEEE Transactions on Multimedia},
  volume = {23},
  pages = {1454--1465},
  publisher = {IEEE}
}

@article{congLinkingBrainResponses2013,
  title = {Linking Brain Responses to Naturalistic Music through Analysis of Ongoing {{EEG}} and Stimulus Features},
  author = {Cong, Fengyu and Alluri, Vinoo and Nandi, Asoke K. and Toiviainen, Petri and Fa, Rui and {Abu-Jamous}, Basel and Gong, Liyun and Craenen, Bart GW and Poikonen, Hanna and Huotilainen, Minna},
  year = {2013},
  journal = {IEEE Transactions on Multimedia},
  volume = {15},
  number = {5},
  pages = {1060--1069},
  publisher = {IEEE}
}

@article{liuEEGbasedStudyPerception2019,
  title = {An {{EEG-based}} Study on Perception of Video Distortion under Various Content Motion Conditions},
  author = {Liu, Xiwen and Tao, Xiaoming and Xu, Mai and Zhan, Yafeng and Lu, Jianhua},
  year = {2019},
  journal = {IEEE Transactions on Multimedia},
  volume = {22},
  number = {4},
  pages = {949--960},
  publisher = {IEEE}
}

@article{song2025recognizing,
  title={Recognizing Natural Images From {EEG} With Language-Guided Contrastive Learning},
  author={Song, Yonghao and Wang, Yijun and He, Huiguang and Gao, Xiaorong},
  journal={IEEE Transactions on Neural Networks and Learning Systems},
  year={2025},
  publisher={IEEE}
}

@inproceedings{songDecodingNaturalImages2024,
  title = {Decoding {{Natural Images}} from {{EEG}} for {{Object Recognition}}},
  author = {Song, Yonghao and Liu, Bingchuan and Li, Xiang and Shi, Nanlin and Wang, Yijun and Gao, Xiaorong},
  booktitle = {International {{Conference}} on {{Learning Representations}}},
  year = {2024},
}

@article{wang2025eegmamba,
  title={EEGMamba: An EEG foundation model with Mamba},
  author={Wang, Jiquan and Zhao, Sha and Luo, Zhiling and Zhou, Yangxuan and Li, Shijian and Pan, Gang},
  journal={Neural Networks},
  pages={107816},
  year={2025},
  publisher={Elsevier}
}

@article{borra2025protocol,
  title={A protocol for trustworthy EEG decoding with neural networks},
  author={Borra, Davide and Magosso, Elisa and Ravanelli, Mirco},
  journal={Neural Networks},
  volume={182},
  pages={106847},
  year={2025},
  publisher={Elsevier}
}


	%Author biography without author photo.


\end{sloppypar}
\end{document}